\documentclass[twocolumn]{aastex631}
\usepackage{amstext,amsmath}

\usepackage[T1]{fontenc}
\usepackage[figure,figure*]{hypcap}
\usepackage{comment}
\usepackage{physics}
\usepackage{soul} 

\usepackage[bb=boondox]{mathalfa}
\usepackage{graphicx}
\usepackage{multirow}
\usepackage{subfig}
\usepackage{colortbl}
\usepackage{float}
\usepackage{enumerate} 

\newcommand{\diff}{\mathop{}\!\mathrm{d}} 
\turnoffeditone
\turnoffedittwo

\begin{document}
\title{\texttt{MEDEA}: A New Model for Emulating Radio Antenna Beam Patterns for 21-cm Cosmology and Antenna Design Studies}

\correspondingauthor{Joshua J. Hibbard}
\author[0000-0002-9377-5133]{Joshua J. Hibbard}
\affiliation{Center for Astrophysics and Space Astronomy, Department of Astrophysical and Planetary Science, University of Colorado Boulder, CO 80309, USA}

\author[0000-0001-5122-9997]{Bang D. Nhan}
\affiliation{Central Development Laboratory, National Radio Astronomy Observatory, Charlottesville, VA 22903, USA}

\author[0000-0003-2196-6675]{David Rapetti}
\affiliation{NASA Ames Research Center, Moffett Field, CA 94035, USA}
\affiliation{Research Institute for Advanced Computer Science, Universities Space Research Association, Washington, DC 20024, USA}
\affiliation{Center for Astrophysics and Space Astronomy, Department of Astrophysical and Planetary Science, University of Colorado Boulder, CO 80309, USA}

\author[0000-0002-4468-2117]{Jack~O.~Burns}
\affiliation{Center for Astrophysics and Space Astronomy, Department of Astrophysical and Planetary Science, University of Colorado Boulder, CO 80309, USA}

\begin{abstract}
    In 21-cm experimental cosmology, accurate characterization of a radio telescope's antenna beam response is essential to measure the 21-cm signal. Computational electromagnetic (CEM) simulations estimate the antenna beam pattern and frequency response by subjecting the EM model to different dependencies, or beam hyper-parameters, such as soil dielectric constant or orientation with the environment. However, it is computationally expensive to search all possible parameter spaces to optimize the antenna design or accurately represent the beam to the level required for use as a systematic model in 21-cm cosmology. We therefore present \texttt{MEDEA}, an emulator which rapidly and accurately generates farfield radiation patterns over a large hyper-parameter space. \texttt{MEDEA} takes a subset of beams simulated by CEM software, spatially decomposes them into coefficients in a complete, linear basis, and then interpolates them to form new beams at arbitrary hyper-parameters. We test \texttt{MEDEA} on an analytical dipole and two numerical beams motivated by upcoming lunar lander missions, and then employ \texttt{MEDEA} as a model to fit mock radio spectrometer data to extract covariances on the input beam hyper-parameters. We find that the interpolated beams have RMS relative errors of at most $10^{-2}$ using 20 input beams or less, and that fits to mock data are able to recover the input beam hyper-parameters when the model and mock derive from the same set of beams. When a systematic bias is introduced into the mock data, extracted beam hyper-parameters exhibit bias, as expected. We propose several future extensions to \texttt{MEDEA} to potentially account for such bias.
\end{abstract}

\section{Introduction} \label{sec:intro}
Experiments and software pipelines designed to extract the extremely weak and highly redshifted cosmological 21-cm spectrum from bright, beam-weighted foregrounds rely heavily on computational electromagnetic (CEM) simulations of the antenna's three-dimensional farfield radiation pattern, or beam pattern. These simulations are then typically used to inform models of systematics, such as the beam-weighted foreground, or to remove systematic spectral structure from experimental data \citep [e.g.,][]{voytek_probing_2014,bernardi_bayesian_2016, bowman_absorption_2018,nhan_assessment_2019,mahesh_validation_2021,kim_impact_2022,spinelli_antenna_2022, monsalve_mapper_2023}. Accurate models of the antenna beam pattern are thus an essential feature of any effort to detect the 21-cm signal. \cite{liu_data_2020} present a more recent review of the impact of different systematics and analyses in the field of 21-cm cosmology.

Full-wave\footnote{Calculations which consist of propagation matrix methods and solutions involving plane wave expansions \citep[e.g.,][]{nagano_numerical_1975,yin_full-wave_2019}, with the latter applying particularly to very low frequency problems.} CEM simulations, such as those generated by CST\footnote{\url{https://www.3ds.com/products/simulia/cst-studio-suite}}, HFSS\footnote{\url{https://www.ansys.com/products/electronics/ansys-hfss}} or FEKO\footnote{\url{https://altair.com/feko/}}, for large antenna structure designs can be computationally expensive and time-consuming, even with the help of high-performance computing (HPC). This is especially true for perturbation studies involving varying different parameters in numerous small increments, such as alignment and dimension variations for different antenna components (e.g., ground screen, dipole length, receiver feed positions), along with the surrounding soil dielectrics~\citep[cf.][]{mahesh_validation_2021,kim_impact_2022}. Hereafter, we shall refer to the physical parameters which affect the antenna beam, such as dipole length or soil dielectric, as beam \textit{hyper-parameters}. Moreover, in-situ measurements of real radio antenna beams provide only scant information of the beam patterns at dynamic ranges of -40~dB \citep{neben_measuring_2015}, probably far above that required for robustly detecting the 21-cm signal \citep{liu_data_2020}. 

Various studies have utilized different decomposition schemes, such as Zernike polynomials fit to antenna holography data  \citep{sekhar_direction-dependent_2022} and spherical harmonics fit to reverberation chamber data \citep{xu_3-d_2017}, to characterize and model the antenna beam patterns. Recently, more advanced approaches have explored using machine learning (ML) models trained on CEM simulations to optimize antenna design parameters \citep[e.g.,][]{akinsolu_machine_2020,linkous_automated_2023,linkous_antennacat_2023}. To our best knowledge, however, currently no model exists for emulating radio beam patterns quickly (on the order of milliseconds) and accurately (for relative beam errors less than -20~dB). Such a model would be of great use not only in 21-cm cosmology to fit for the beam patterns in sampling algorithms, but also for general-purpose antenna design studies, where it could be used to quickly generate beam patterns resulting from small changes in antenna configurations or dielectric properties, facilitating rapid prototyping.

In this study, we present a rapid beam pattern simulation emulator algorithm\footnote{\url{https://github.com/CU-NESS/medea}, and Zenodo 
\url{https://zenodo.org/doi/10.5281/zenodo.11584632}}, the Model for Emulating Directivities and Electric fields of Antennas (\texttt{MEDEA} \footnote{See for instance, Hesiod's \textit{Theogony}, 956-962, or \textit{Medea} by Euripides.}). This emulator is not meant to capture all features of the ground truth radio beam pattern of an actual experiment; note, however, that such is also true for the full-wave numerical CEM simulations. Rather, the emulated beams are used as a starting point for the model of the true antenna beam pattern, as is done in all 21-cm studies of global signal extraction and beam pattern estimation. The emulator, capable of representing thousands of simulations, thus presents a more exhaustive starting point for the model.

To use \texttt{MEDEA}, the emulator must be provided an input set of beams---either analytical or numerical---which have been generated on a low-resolution grid, typically on the order of ten beam hyper-parameters. To then test the emulator's accuracy for this grid resolution, it is necessary to remove one of the input beams from the input set as the test beam, use \texttt{MEDEA} to generate the missing test beam from interpolation, and then compare the interpolated beam to the true, test beam. If smaller beam errors are required, then the resolution of the input set of beams is increased. If the input set consists of a regular, rectangular grid, then the resolution must be effectively doubled, and the test of accuracy repeated until the desired level of beam error is attained.

The \texttt{MEDEA} emulator consists of two main parts. Firstly, a set of three-dimensional farfield beam patterns over a coarsely sampled hyper-parameter space is computed with either conventional full-wave CEM programs (the Microwave Suite CST was used in this study), or closed-form analytical solutions. Some of these hyper-parameters include the soil (or regolith\footnote{In this paper, soil and regolith are interchangeable.}) dielectric constant (also known as relative permittivity), $\epsilon$, as well as the dipole arm length, $l$. For example, in \cite{mahesh_validation_2021} and \cite{monsalve_mapper_2023}, beam patterns are generated for different soil dielectrics and ground screen sizes, while \cite{kim_impact_2022} simulate different receiver feed alignment offsets relative to the HERA's 14-meter dish's vertex. These simulations form the \textit{input} set which defines the basic shapes of the antenna's beam pattern and its dependence on the beam hyper-parameters. Secondly, this input set is decomposed into coefficients in a complete, linear basis which fully describes a subset of the $4\pi$ steradians of the sky. The coefficients are then interpolated to form new values between the coefficients of the sampled input beam set. Subsequently, a new beam pattern is reconstructed using the interpolated coefficients and the linear basis.

We study \texttt{MEDEA}'s speed and accuracy in emulating realistic antenna beam patterns with three antenna cases. The first one is an analytical calculation of a small, horizontal dipole suspended at a height, $h$, above an infinite perfect electric conductor (PEC) ground plane. This simple model exhibits the generic electromagnetic characteristics, namely chromaticity, azimuthal asymmetry, and sidelobes which depend on the hyper-parameter $h$. Additionally, two numerically simulated antenna cases are studied in this work: the first is a patch antenna attached to a spacecraft which is on the lunar regolith with dielectric constant $\epsilon$ and observing in the Cosmic Dawn band ($\sim 50-100$~MHz), and the second case is inspired by the recent landed ROLSES radio instrument \citep{burns_low_2021} on the NASA Commercial Lunar Payload Services (CLPS) program and consists of a pair of dipole Stacer antennas attached to a lander which is also situated on the lunar surface but observing at lower frequencies in the Dark Ages band ($\sim 1-30$~MHz). 

To produce the second part of the model, namely the complete, linear basis describing a subset of the sky, we combine a technique used in galaxy masking with the experimental setup common in 21-cm cosmology and ground-based radio telescopes. That is, we assume that the portion of the sky to be masked is the observer horizon surrounding the antenna, for which the masking and harmonic analysis code \texttt{Cryofunk} \citep{gebhardt_harmonic_2022} is ideal. This code produces arbitrary spatial masks of the $4\pi$ steradian sky and then calculates a linear basis (whose vectors are orthogonal and linear combinations of spherical harmonics) for the unmasked portion. Because this basis is complete, \textit{it can describe any function, and hence any beam pattern, defined within the boundaries of the mask}. The benefit of decomposing only the sky above the horizon means that each degree of freedom (DoF) provided by a \texttt{Cryofunk} mode (weighted by a Cryo\footnote{Cryo = \underline{Cr}eate \underline{y}our \underline{o}wn}-coefficient, as used hereafter) contributes solely to the DOFs in the radio antenna beam pattern, whereas decompositions of the entire $4\pi$ sky using traditional spherical harmonics are forced to incorporate the entire horizon in the spherical harmonic coefficients.

The \texttt{Cryofunk} basis then depends only on the topological shape of the horizon along with the spatial resolution of the sky. Hence, it can be used for any beam pattern with given hyper-parameters at all frequencies in the antenna's bandwidth. Hereafter, beam patterns produced by \texttt{MEDEA} are denoted as \textit{Cryo-beams} and the fiducial simulated or analytically calculated beams in the input set are denoted for brevity as \textit{CST-beams}. It will be clear from context, or explicitly labeled, when the latter beams are either produced by CEM software or calculated analytically.

In the second part of this work, we test the ability of \texttt{MEDEA} as a model to fit radio spectrometer data, and in particular, spectral data typically studied in global 21-cm cosmology experiments. We subsequently generate noisy, mock spectra using a galactic foreground weighted with a representative antenna beam for each of the antenna cases, add thermal noise, and then form a likelihood with diagonal noise covariances. The model used in the likelihood consists of a beam topological pattern generated by \texttt{MEDEA} in combination with a common galactic foreground model. Using the nonlinear sampling code \texttt{Polychord}, we then fit for both the galactic foreground spectral index parameters and the beam hyper-parameter, in order to determine the degree to which the latter can be extracted from radio spectrometer data consisting solely of a radio antenna beam, the galactic foreground, and thermal noise.

The rest of this work is structured as follows. In Section~\ref{sec:methods-beam-em}, we introduce the linear beam model \texttt{MEDEA}, the three antenna case studies, and the interpolation methods used for emulation, including splines and Gaussian Process Regression (GPR). Then, in Section~\ref{sec:methods-fitting}, \texttt{MEDEA} is also tested as a model with foreground and beam parameters in a nonlinear sampler \citep[\texttt{Polychord},][]{handley_polychord_2015} to determine if a fiducial beam's hyper-parameters can be extracted by fits to mock beam-weighted foreground spectra. We also study the extraction of the beam hyper-parameters when a systematic bias is introduced into the mock spectrum. In Section~\ref{sec:results-beam-em}, we first examine the accuracy of the interpolation methods by comparing the interpolated to the corresponding values of the input set, and in Section~\ref{sec:results-fitting} we then examine the errors on hyper-parameters extracted from spectra using a nonlinear sampler. We discuss possible model extensions of \texttt{MEDEA} in Section~\ref{sec:discussion} and conclude in Section~\ref{sec:conclusions}.

\section{Beam Emulation}
\label{sec:methods-beam-em}
Since the \texttt{Cryofunk} code works in the space of \texttt{Healpix} maps \citep{gorski_healpix_2005}, we convert all three-dimensional farfield patterns used in this study from their original spherical coordinate grids (of $1^{\circ}\times1^{\circ}$ in $\theta$ and $\phi$) to \texttt{Healpix} format using a rectangular bivariate spline. Furthermore, to maintain the high numerical precision for our analysis from the CST farfield solutions, we employ a custom Visual Basic (VBA) script to export the Farfield Field Source (FFS) CST data using the argument \texttt{ASCIIExportAsSource} for the built in CST function \texttt{FarfieldPlot} instead of the usual farfield export. Without the FFS setting, the exported farfield data contain significant rounding errors. Lastly, we note that all analyses in this work use \texttt{Healpix} maps with parameter $N_\mathrm{side} = 32$, although all results apply to higher resolution maps with the only difference being that more Cryo-coefficients must be interpolated, as the number of basis vectors scales as the number of pixels (because the basis is complete, the number of basis vectors is equal to the number of pixels above the horizon in the map). This $N_\mathrm{side}$ parameter corresponds approximately to an angular resolution of $1.83^{\circ}$. Applications requiring greater angular resolution can use a higher $N_\mathrm{side}$ parameter, although this then increases the number of basis vectors required to fully describe the sky. If on the other hand the beam pattern is highly directive with only one main lobe, all but the latter could potentially be masked, lowering the number of required basis vectors. For this work we have chosen $N_\mathrm{side} = 32$ throughout as such scales are typical of the large beams used in global 21-cm cosmology.

\subsection{Linear Beam Model - Cryobeam}
Working in \texttt{Healpix} where a pixel is denoted by an index $i$, the portion of the beam gain, directivity, or radiation intensity pattern (hereafter referred to simply as the beam pattern), $B_i$, not blocked by the horizon $H_i$ can be decomposed in \texttt{Cryofunk} space \citep{gebhardt_harmonic_2022} as 
\begin{equation}
    B_i(\nu, \boldsymbol{\zeta})|_{i \rightarrow N_H} = \sum_l^{N_b} Y_{il} k_l(\nu,\boldsymbol{\zeta}),
    \label{eqn:beam-definition}
\end{equation}
where $Y_{il}$ are the \texttt{Cryofunk} basis functions, which are linear combinations of spherical harmonics and form a complete, orthogonal basis, and $k_l(\nu,\boldsymbol{\zeta})$ are the coefficients in this basis. $N_H$ and $N_b$ denote the number of pixels above the horizon and the number of basis vectors, respectively. The vector $\boldsymbol{\zeta}$ denotes the beam \textit{hyper-parameters} which control the topological shape of the farfield beam pattern, such as the soil dielectric constant $\epsilon$ or the height of the antenna above the ground $h$. In principle, any beam hyper-parameter can be studied and included in this framework, although if the beam pattern changes chaotically or exhibits rapid nonlinear dependence as a function of $\boldsymbol{\zeta}$, the accuracy of interpolation will decrease. See Section~\ref{sec:results-beam-em} for a discussion and visualization on this.

Note again that we only decompose the unmasked portion of the beam patterns above the observer horizon (i.e., $i$ runs up to pixel $N_H < N_{pix}$, where the latter is the total number of pixels in the \texttt{Healpix} map). Furthermore, the basis functions $Y_{il}$ only depend upon the shape of the horizon $H_i$ and are thus the same across the observing band $\nu$ and for every beam hyper-parameter in $\boldsymbol{\zeta}$.

Typically $N_b$, the number of basis vectors, is given by\footnote{However, it is possible to use a smaller number of basis vectors, depending on the accuracy required. See Section~\ref{sec:discussion} for a discussion on truncating the number of modes.} $N_b = N_{pix} - N_H$. We can thus use Equation~\eqref{eqn:beam-definition} to write
\begin{equation}
\begin{aligned}
    \sum_i^{N_{pix}} B_i(\nu, \boldsymbol{\zeta}) H_i = \sum_l^{N_b} \sum_i^{N_H} Y_{il}k_l(\nu, \boldsymbol{\zeta}).    
\end{aligned}
\label{eqn:base-cryo-beam}
\end{equation}
Equation~\eqref{eqn:base-cryo-beam} represents the final linear model for the beam pattern given both a horizon profile $H_i$ and a beam set $B_i(\nu, \boldsymbol{\zeta})$ generated using multiple values of beam hyper-parameters.

\subsection{Antenna Case Studies}
\label{sec:antenna_cases}
We study three different antenna cases in order to understand how \texttt{MEDEA} can be used to rapidly generate accurate beam patterns, as well as to understand its potential limitations. 

\subsubsection{Analytical Dipole}
\label{sec:analytical-dipole}
The first case study is an analytically tractable, small horizontal dipole suspended a height $h$ above an infinite PEC conductor. As it exhibits both chromaticity (frequency-dependent beam patterns) and complex dependence upon its hyper-parameter $h$ (including effects on the number and shape of sidelobes, for instance), this well-studied beam pattern illustrates how \texttt{Cryofunk} decomposes complex beam patterns and serves as a baseline for comparison with the subsequent full-wave numerical beam simulations.

Its radiation intensity pattern (the non-normalized directivity) is given by the following equation \citep[][eq. 4-121, pg. 203]{balanis_antenna_2005}:
\begin{equation}
    B_d(h,\theta,\phi) = A (1 - \sin^2{\theta} \sin^2{\phi}) \sin^2{(kh \cos{\theta})}
    \label{eqn:dipolegroundplane},
\end{equation}
where $k$ is the wave number and $A$ is a normalizing factor accounting for the impedance of the medium, the length of the dipole, and the current. We plot several representative beams patterns as polar plots in the top panel of Figure~\ref{fig:polar-plots}, where each color denotes a different value of the beam hyper-parameter $h$. We simulate this beam in the $50 - 100$~MHz band, with $h$ ranging from 1 to 3~meters, and test hyper-parameter grid resolutions of $\Delta h = [0.1, 0.05, 0.025$~m] corresponding to 11, 21, and 41 simulated beams in the input set, respectively.

\subsubsection{Numerically Simulated Antennas}
\label{sec:numerical}
For a more realistic comparison, we study two antennas generated from numerical full-wave simulations using the time-domain solver in CST. Both antennas are inspired by instruments for lunar radio experiments, though they are similar enough to a number of ground-based dipole antennas for 21-cm experiments. 

\begin{figure}[t!]
    \centering
    \includegraphics[width=1\columnwidth]{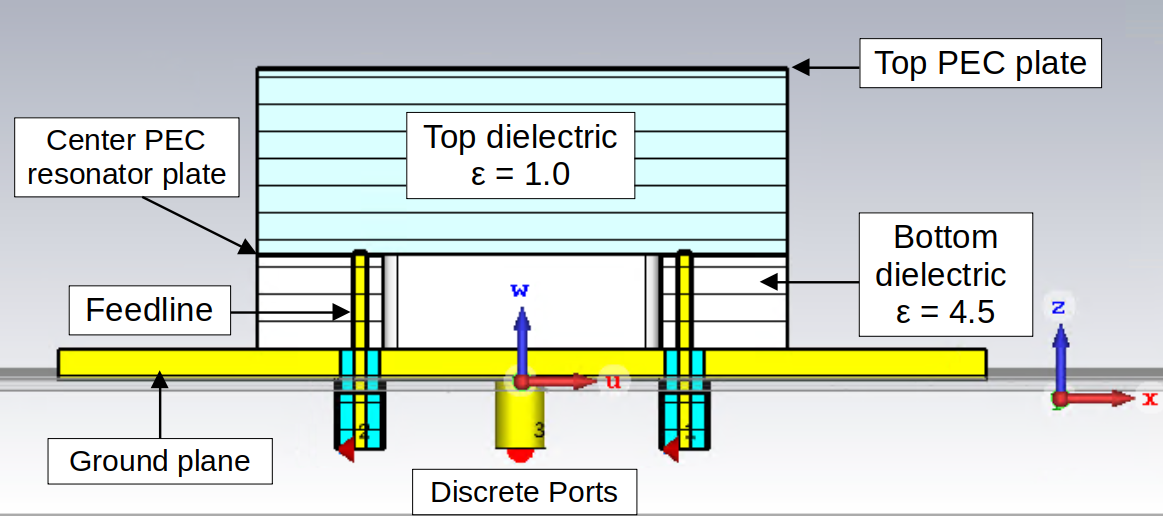}
    \includegraphics[width=1\columnwidth]{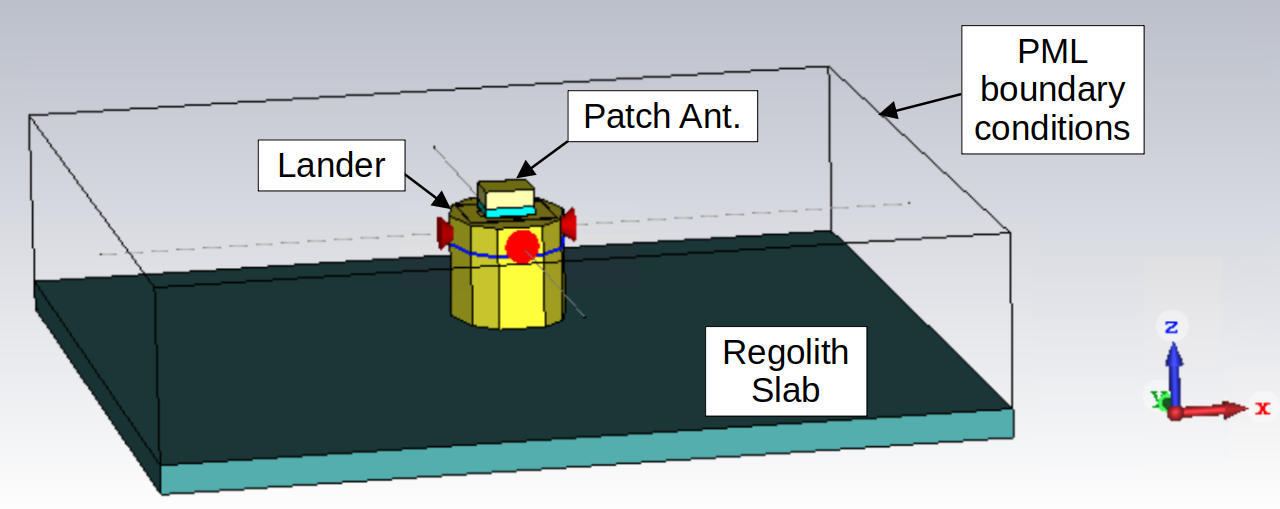}    
    \caption{Rendering of the Patch Antenna simulation setup in CST using the FIT solver and contained within a bounding box with PML boundary conditions. The top panel shows the cross Section~and interior construction of the Patch Antenna. The bottom panel shows the location of the Patch Antenna situated on top of a simplified generic lander resting on a lunar regolith slab.}
    \label{fig:cst_patch_ant_render}
\end{figure}

The first numerical beam is a patch type antenna situated on top of a simplified lunar lander (denoted hereafter as the Patch Antenna), originally designed in an attempt to achieve a relatively smooth frequency response across the $40-120$~MHz band, making it useful for 21-cm cosmology global signal experiments. The patch antenna itself consists of two layers of dielectric materials sandwiched between a thin square PEC plate (0.85~m~$\times$~0.85~m~$\times$~1~mm) on the top, a thin PEC resonator plate in the center, and a square PEC ground plate at the bottom which is overlapped by the top of the lander. The center resonator and bottom dielectric layer ($\epsilon=4.5$, with a thickness of 0.15~m) are excited by four individual discrete ports with a port impedance of $50~\Omega$. To form a dipole-like beam pattern, opposite-facing pairs of antenna ports are simultaneously excited with a stimulation signal $180^{\circ}$ out of phase. The upper dielectric layer ($\epsilon=1.0$, with a thickness of 0.3~m) and the PEC plate help with the impedance tuning in order to broaden the operational bandwidth. Renderings of the Patch Antenna model are shown in Figure~\ref{fig:cst_patch_ant_render}.

The lander is modeled as an elongated hexagonal cylinder with dimensions of 2.6~m~$\times$~2.0~m~$\times$~2.4~m and made of PEC material, all resting on top of the regolith slab. The regolith underneath is simulated as an 8~m~$\times$~8~m square slab with a depth of 0.5~m and uses the perfectly matched layer (PML) boundary condition method to approximate an infinite regolith interface using CST's Finite Integration Technique (FIT) time-domain solver\footnote{Due to the much faster simulation time achieved by the FIT time-domain solver for all frequencies of interest at once, we only spot checked our FIT results with CST's Method of Moments (MoM) frequency-domain solver at a few frequencies for a single $\epsilon$ value and they are in general agreement. Since we were not trying to compare our models with a ground-truth antenna beam in this study, self-consistency between simulation realizations within the same FIT solver is what matters. Thus we ensure all simulation settings are identical for our two numerical antenna models.} \citep{davidson_computational_2010}. Based on lunar regolith electromagnetic property estimates from previous studies \citep[e.g.,][]{olhoeft_dielectric_1975,heiken_book-review_1991,li_lunar_2022}, we simulated representative ranges of the dielectric constant, $\epsilon = 1.5 - 4.5$ with increments of $\Delta\epsilon = 0.25$, and the loss tangent $\tan\delta = 0.001 - 0.02$, with increments of $\Delta\tan\delta =0.0025$. In this analysis, however, we find that the variations in $\epsilon$ have a dominant effect on the farfield pattern compared to those of $\tan\delta$; hence, for our current purpose, we only use Patch Antenna beam simulations with a fixed value of $\tan\delta = 0.005$.

\begin{figure}[t!]
    \centering    
    \includegraphics[width=1\columnwidth]{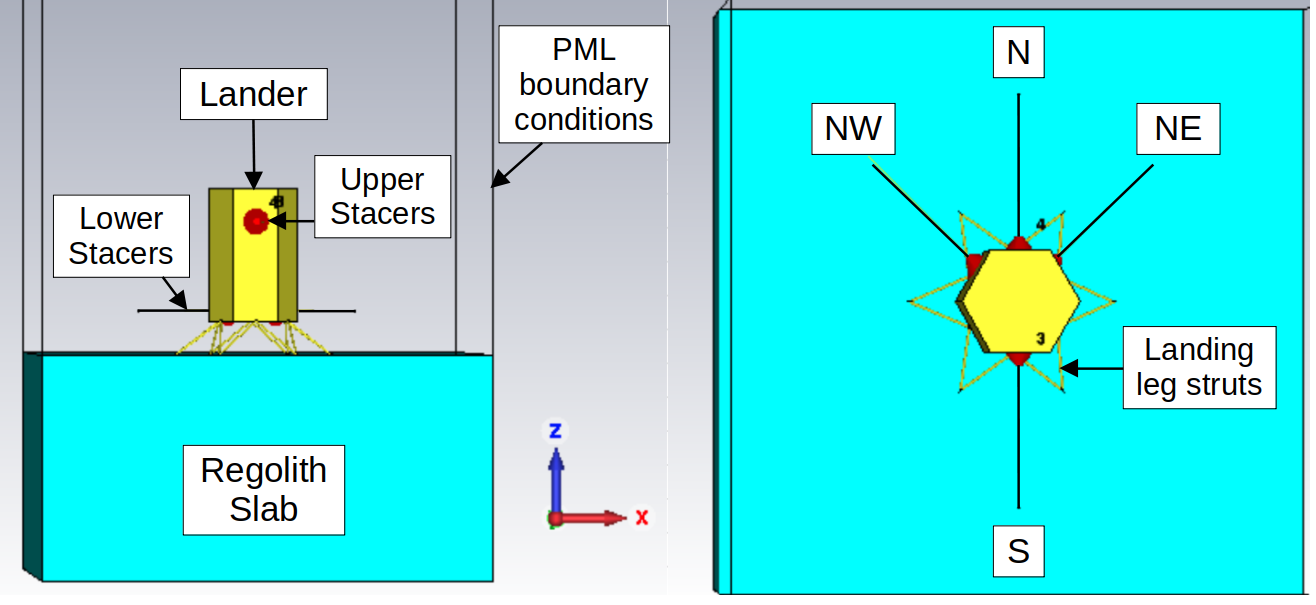}    
    \caption{Rendering of the ROLSES-inspired Stacer antennas simulation setup in CST using the same FIT solver and PML boundary conditions as the Patch Antenna. Depicted are the side view (left) and top view (right) of the simplified ROLSES lander above the lunar regolith on landing struts.}
    \label{fig:cst_rolses_ant_render}
\end{figure}

The second antenna model is based upon the ROLSES radio telescope, introduced above \citep{burns_low_2021}. The simplified antenna model studied here assumes four Stacer (hollow tube telescoping) monopole antennas of length 2.5~m extended horizontally from the faces of a PEC hexagonal cylindrical lander with extended landing leg struts (see Figure~\ref{fig:cst_rolses_ant_render}). This simplified lander model is 2.0~m~$\times$~2.0~m~$\times$~3.0~m and 0.75~m above a lunar regolith slab which is set to have a side of 12~m and a depth of 4~m, and similar PML boundary conditions as used for the Patch Antenna. The regolith slab is also assumed to have a single composite relative dielectric constant $\epsilon$ and loss tangent $\tan\delta$. The antennas were simulated for the bandwidth of $1-30$~MHz. 

ROLSES consists of two pairs of Stacer monopoles. The upper pair is located on the North and South panels at a height of 2.7~m above the regolith, while the lower pair is located on the North East (NE) and North West (NW) panels at a height of 1.0~m above the regolith. To create a dipole pair, the upper Stacers placed at $180^{\circ}$ apart are phased together to form a dipole beam pattern, for different lunar regolith dielectric constants. The latter have a range of $\epsilon = 5 - 6$ with a finer increment of $\Delta\epsilon = 0.025$ than for the Patch Antenna for a more refined study on the effects of low-level variations in the regolith's dielectric on the farfield beam patterns, and also a different fixed value of $\tan\delta = 0.0159$. See Appendix \ref{app-rolses-dipole} for a derivation of the dipole beam pattern. 
Renderings of the antenna model are shown in Figure~\ref{fig:cst_rolses_ant_render}.

\begin{figure*}[htb!]
    \centering
    \includegraphics[width=0.96\textwidth]{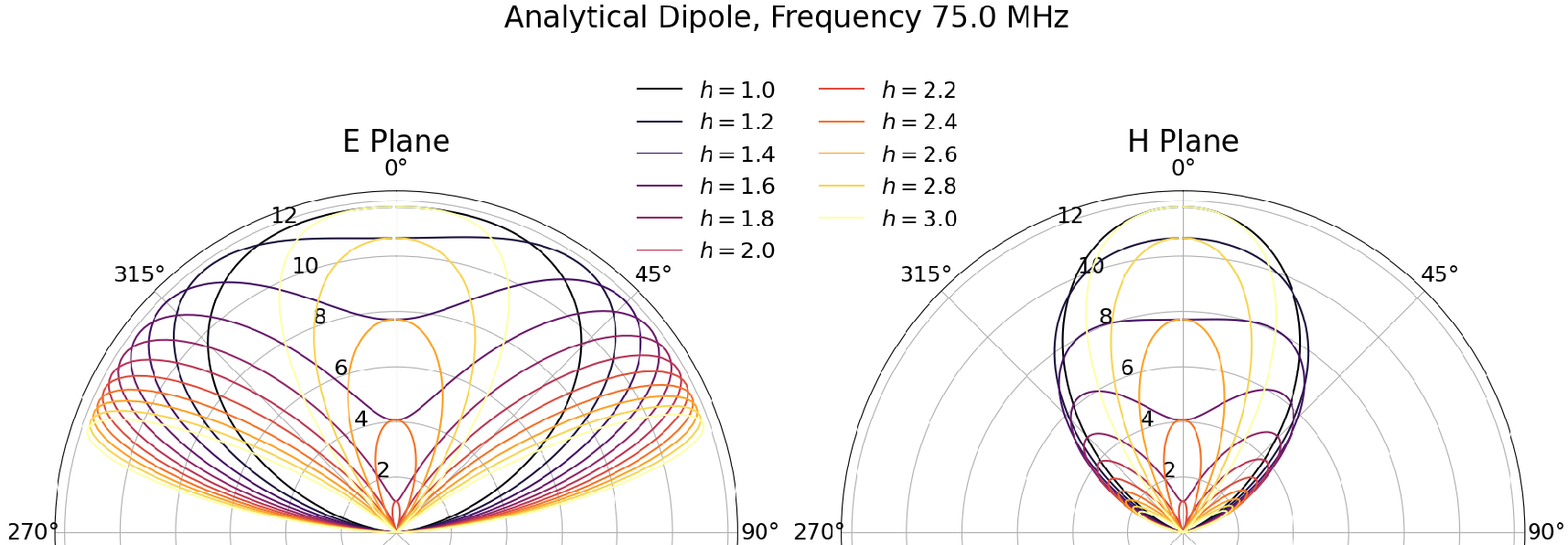}
    \includegraphics[width=0.96\textwidth]{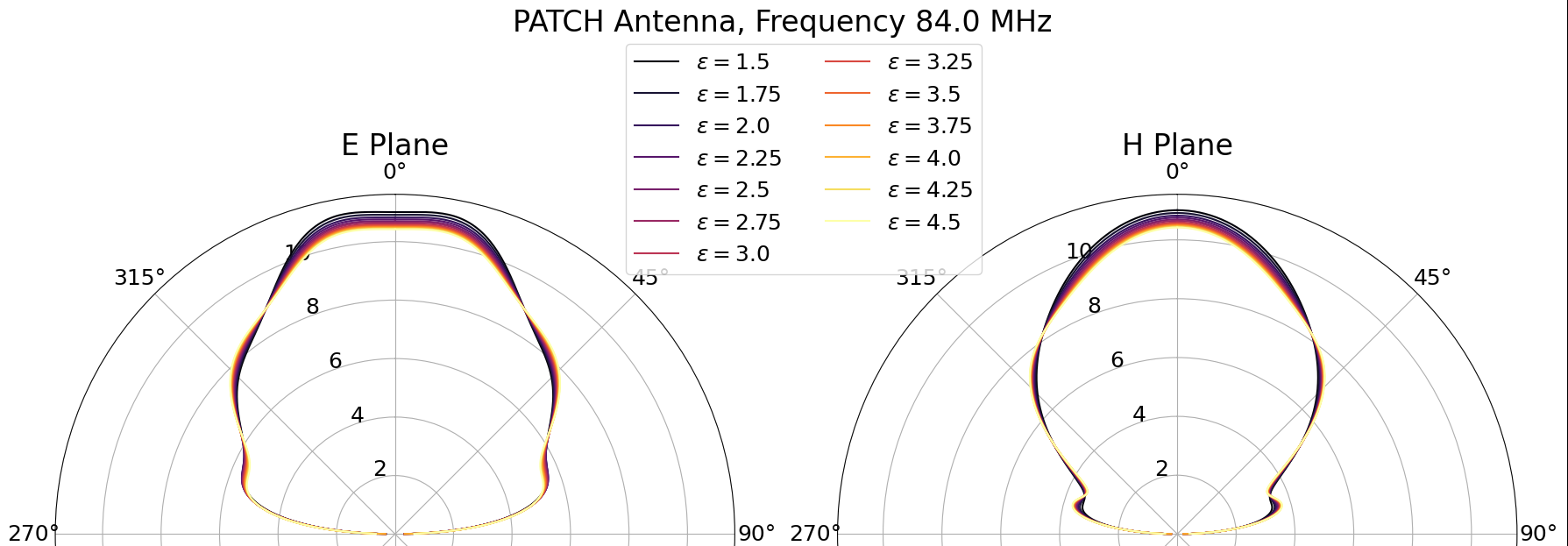}
    \includegraphics[width=0.96\textwidth]{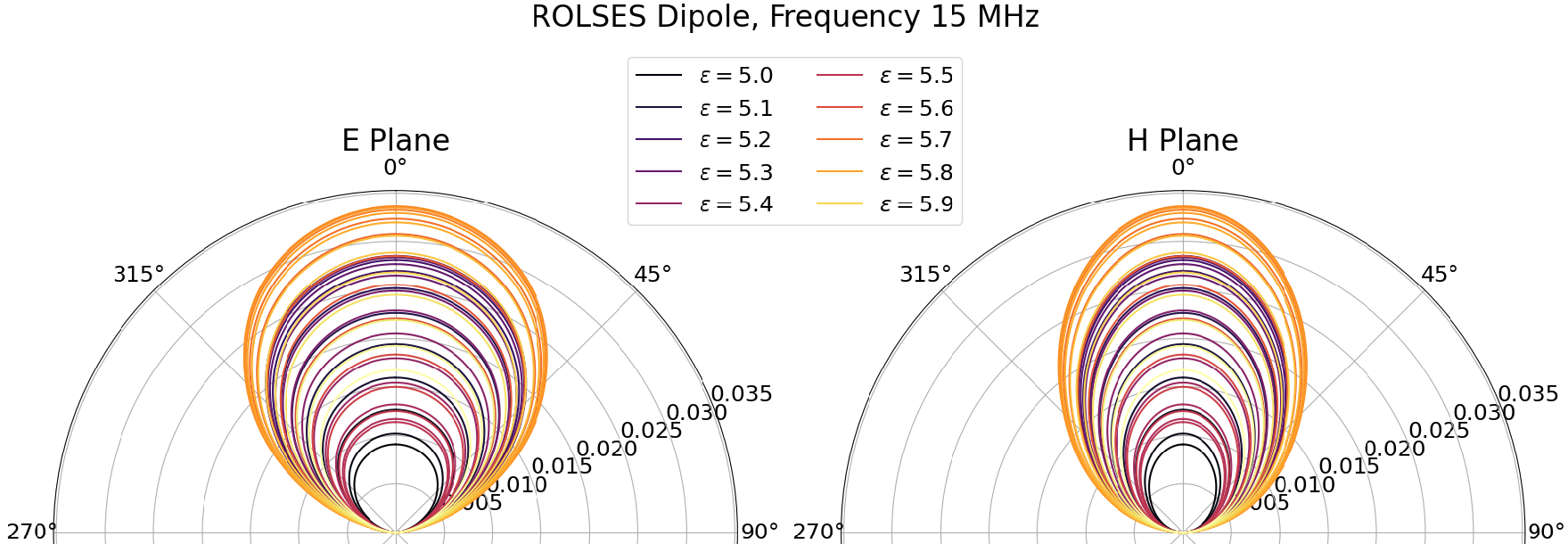}
    \caption{Polar plots of the non-normalized linear gains for the analytically and numerically simulated beam patterns. The Analytical Dipole at its mid-band frequency is shown in the top panel as a function of hyper-parameter $h$. The middle panel shows the Patch Antenna at 84~MHz and the bottom shows the ROLSES Dipole at 15~MHz. Both of the latter two beams are simulated numerically in CST for different values of the lunar regolith dielectric constant, $\epsilon$, with darker colors denoting lower values and lighter colors higher values. The left column of each panel shows the E plane cut, while the right column shows the H plane cut. Note that the ROLSES Dipole is highly nonlinear in dielectric space, due to the fact that the antenna design is not optimized for the $1-30$~MHz band. We only show the bottom half ($\theta \leq 90 ^{\circ})$ of the polar plots.}
    \label{fig:polar-plots}
\end{figure*}

The ROLSES Dipole antenna model is representative of an experiment which is neither optimized for 21-cm cosmology nor for observing in the $1-30$~MHz band, given the length of the Stacer antennas. This antenna model is thus used here as a \textit{sub-optimal} case, where the beam pattern is not only complex spatially, but exhibits a highly nonlinear behavior as a function of the beam hyper-parameter $\epsilon$. As shown in the bottom panel of Figure~\ref{fig:polar-plots}, the polar plots for the ROLSES Dipole oscillate rapidly across the small regolith dielectric range simulated. As we shall see, its coefficients from \texttt{Cryofunk} reflect this rampant nonlinearity. The ROLSES Dipole though is useful here to both be contrasted with our optimal case, the Patch Antenna, and probe the limits of \texttt{MEDEA}'s accuracy for sub-optimized antennas.

Note that the ROLSES Dipole is also simulated at hyper-parameter grid resolutions of $\Delta \epsilon = [0.1, 0.05, 0.025]$ corresponding to 11, 21, and 41 beams. On the other hand, the Patch Antenna is simulated at only two resolutions, $\Delta \epsilon = [1, 0.5]$, corresponding to three and seven beams. These choices of dielectric grid resolutions were driven primarily by the computational cost of these simulations.~The Patch Antenna, in particular, could have been simulated at a finer resolution for higher interpolation accuracy, although it appears to be likely unnecessary given the linear behavior of the Cryo-coefficients (see below). For this work, in fact, it suffices for us to demonstrate the accuracy and efficiency of \texttt{MEDEA} as a beam model for the two chosen patterns.

Polar plots of the E and H plane projections at midband frequencies for each numerical antenna case are shown in Figure~\ref{fig:polar-plots} for all regolith dielectric constants simulated. For a linearly polarized antenna aligned in the $x$-axis in the Cartesian coordinate, the $E$-plane corresponds to the principle plane containing the electric field vector (given here by the $\phi = 0^{\circ}$ plane) while the $H$-plane is perpendicular to the former and contains the magnetizing field vector (and is given by the $\phi = 90^{\circ}$ plane). We note that the Patch Antenna beam pattern is relatively linear with respect to the dielectric constant, at least compared to the other two antenna case studies. That is, it does not exhibit oscillations which cross previous beam patterns for a monotonically increasing hyper-parameter (compare with the ROLSES Dipole and the Analytical Dipole). 

Lastly, in both the analytical and farfield simulation beam cases, we test how normalizing the beams before decomposition into Cryo-coefficients affects subsequent coefficient interpolation. We normalize by dividing the beam pattern by its $4\pi$ steradian integral at each frequency $\nu$ and at every hyper-parameter in $\boldsymbol{\zeta}$. That is, we enforce
\begin{equation}
    \widetilde{B}(\Omega,\nu,\boldsymbol{\zeta}) = \frac{B(\Omega, \nu, \boldsymbol{\zeta})}{\int_{4\pi} \diff\Omega B(\Omega, \nu, \boldsymbol{\zeta}) }\,, 
    \label{eqn:norm}
\end{equation}
where the tilde denotes the normalized beam. We also normalize the beams in this manner when comparing the different antenna cases.

\subsection{Direct Decomposition}
Armed with our three antenna cases, we begin by decomposing each of them directly into their respective coefficients from \texttt{Cryofunk}. This will allow us to examine the behavior of the coefficients as a function of hyper-parameter and determine the limiting, numerical precision of \texttt{MEDEA}. That is, we compute
\begin{equation}
    k_l(\nu, \boldsymbol{\zeta}) = Y_{il}^{-1} B_i(\nu, \boldsymbol{\zeta})
\end{equation}
to acquire the Cryo-coefficients, and then use them to reform the subsequent Cryobeam via $B_i = Y_{il} k_l$.

\begin{figure}[htb!]
    \centering
    \includegraphics[width=0.46\textwidth]{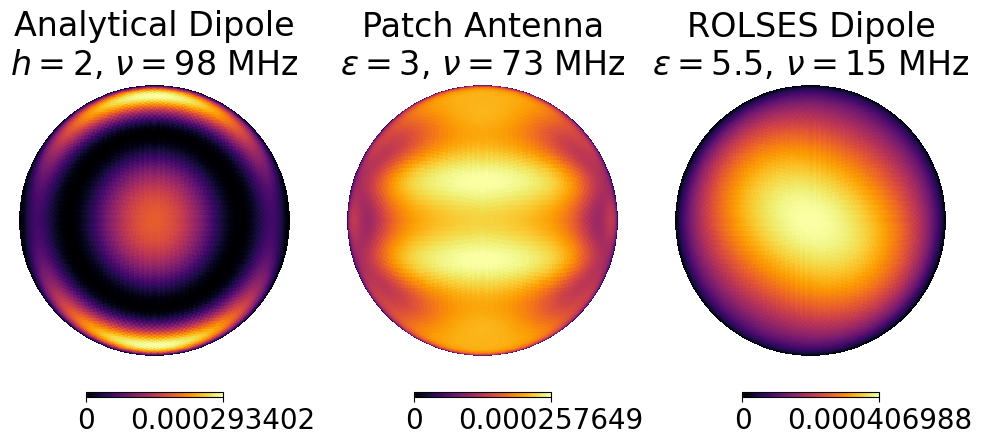}
    \caption{Orthographic projection of the direct decomposition of all antenna cases. Each antenna case is shown at a representative beam hyper-parameter and mid-band frequency. We note that these beams exhibit significant topographical complexity, but that \texttt{Cryofunk} has little difficulty in describing these complex beam patterns up to numerical precision in floating-point operations.}
    \label{fig:direct-decomposition}
\end{figure}

Several example Cryobeams from direct decomposition along with their CST counterparts are shown in Figure~\ref{fig:direct-decomposition} for reference: the Analytical Dipole (left), the Patch Antenna (middle), and the ROLSES Dipole (right). Each antenna case is shown at a representative beam hyper-parameter and mid-band frequency, as indicated in each subtitle. Note that the antenna gains shown in each panel are normalized as described in Equation~\eqref{eqn:norm} to facilitate comparison between different beams.

\subsection{Beam Interpolation Methods and Cryo-coefficients}
\label{sec:methods-beam-interp}

\begin{figure}
    \centering
    \includegraphics[width=1\columnwidth]{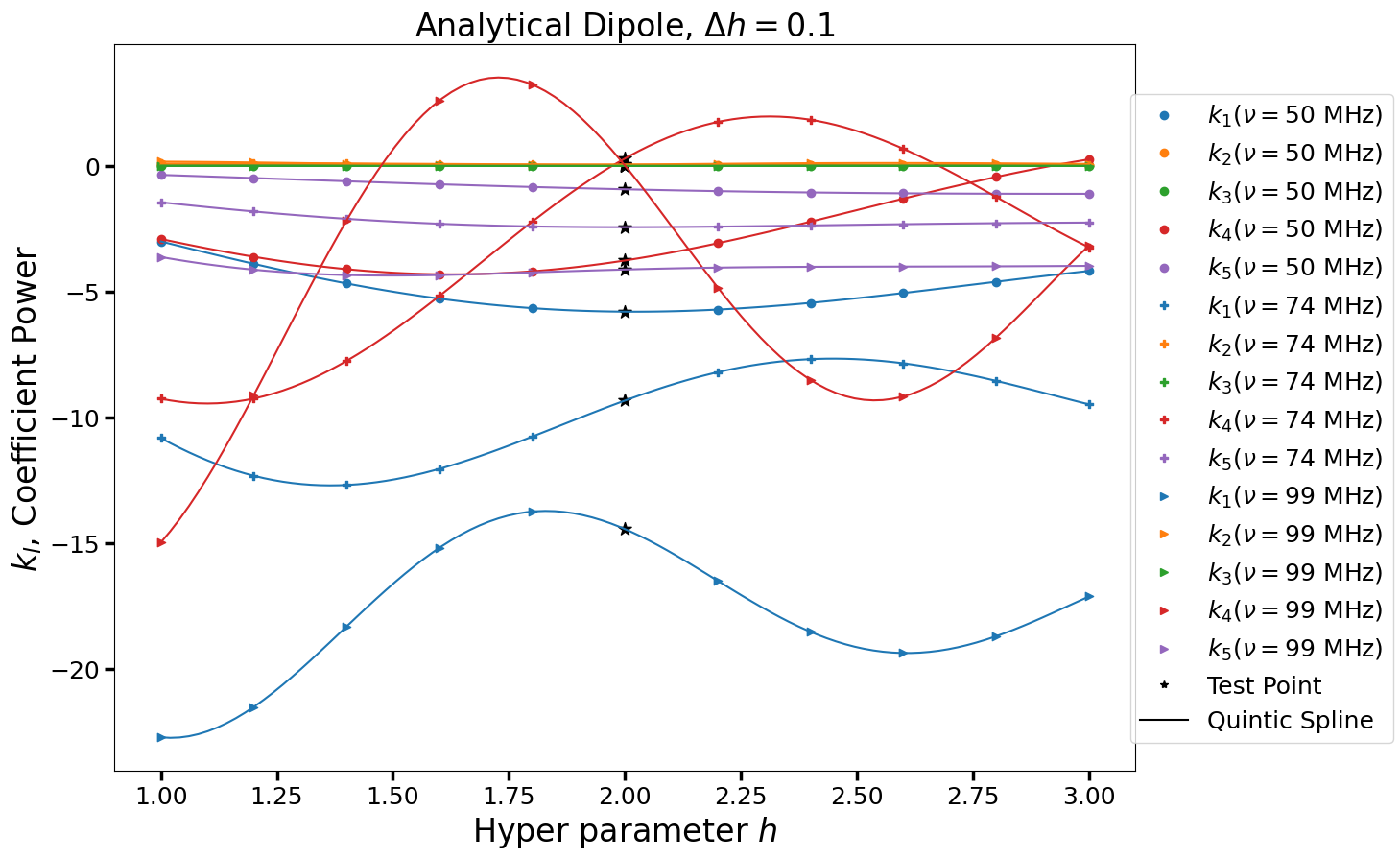}
    \includegraphics[width=1\columnwidth]{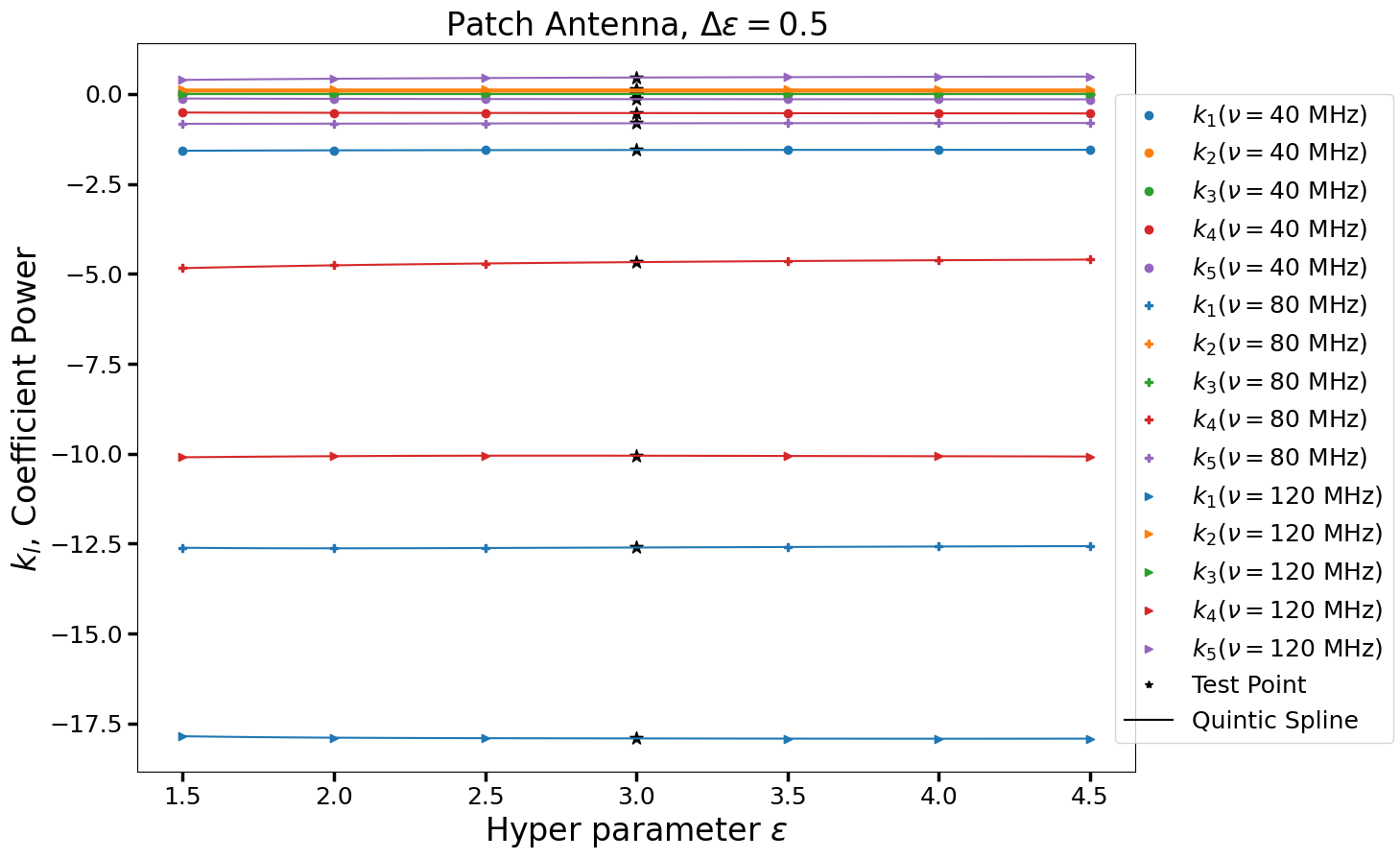}
    \includegraphics[width=1\columnwidth]{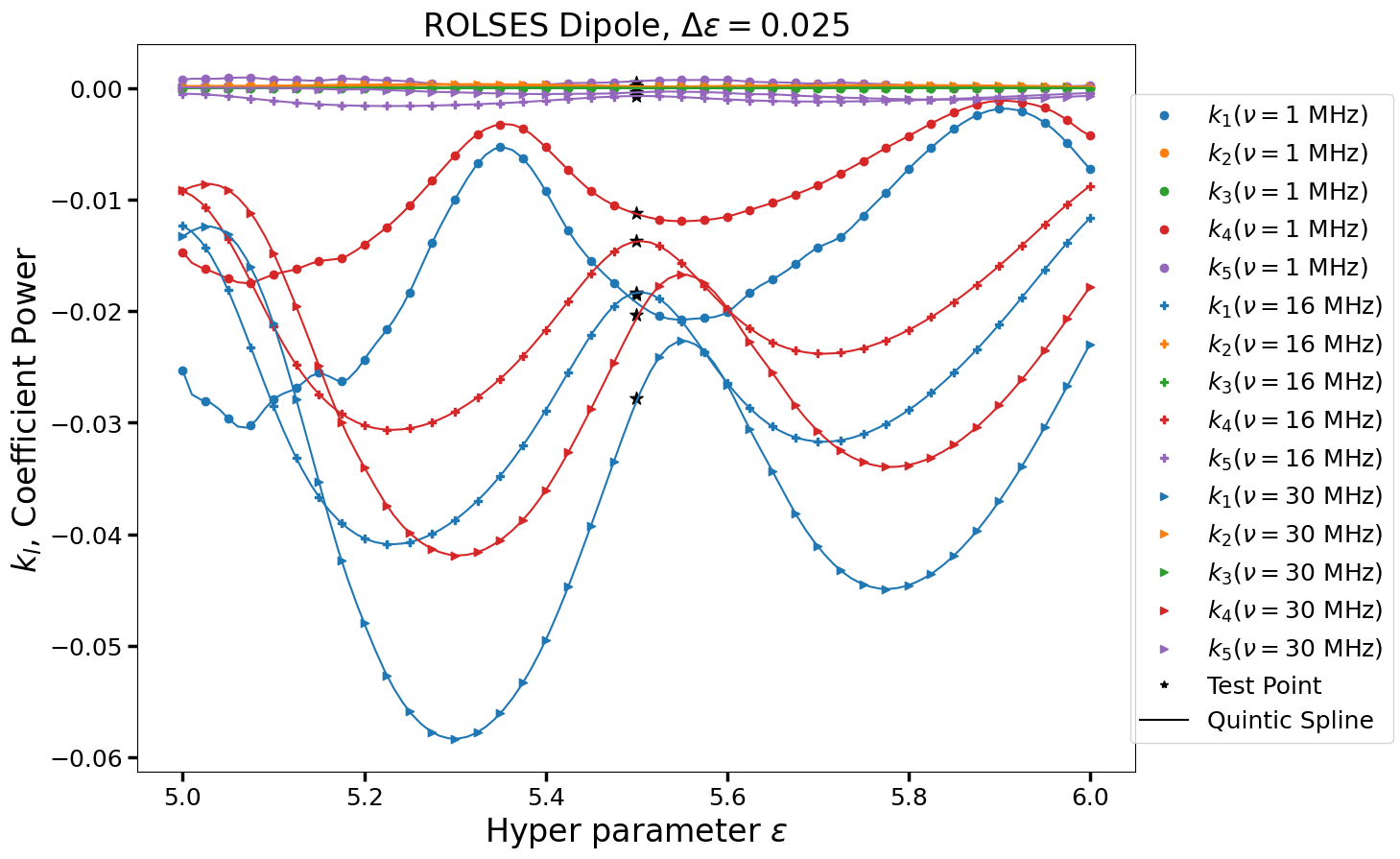}
    \caption{The first five Cryo-coefficients' power plotted as a function of beam hyper-parameter for each antenna case study and at three representative frequencies. The different colors label the different coefficients $k_l$, while the circles represent low-band, the plus signs mid-band, and the triangles high-band frequency beams; the exact frequencies represented are shown in the legend for each antenna case. The black stars correspond to the points in the set $K_T$ that are tested for interpolation, while the solid lines are quintic splines fit to the input sets $K_{I-1}$ which do \textit{not} contain the test set of coefficients.}
    \label{fig:cryo-coeff}
\end{figure}

For a given horizon profile, the beam pattern structure is stored entirely in the magnitudes of the Cryo-coefficients\footnote{Throughout the rest of this work, we shall occasionally refer to the Cryo-coefficients simply as the \textit{coefficients} when it is clear from context.} $k_l$. In Figure~\ref{fig:cryo-coeff}, we plot the first five Cryo-coefficients at three representative frequencies as a function of hyper-parameter for each antenna case. Each color represents a different coefficient $k_l$ (with units of gain for coefficient power), while each symbol denotes the frequency of the beam from which the coefficients derive within each antenna's band: circle for a low-band frequency beam, plus-sign for the mid-band, and triangle for the high-band. The exact frequencies represented are also noted in the legend of Figure~\ref{fig:cryo-coeff}. As in the previous figure, the top panel shows the Analytical Dipole, the middle panel the Patch Antenna, and the bottom panel the ROLSES Dipole.

The Analytical Dipole and Patch Antenna coefficients are relatively smooth as a function of beam hyper-parameter, with the latter being nearly linear for all coefficients and frequencies. In contrast, the ROLSES Dipole exhibits smoothness in some regions of the parameter space, notably between roughly $\epsilon = 5.2 - 5.8$, but is relatively irregular below $\epsilon = 5.2$, as shown by the blue circles corresponding to the first Cryo-coefficients. Indeed, the low-band ROLSES Dipole coefficients fluctuate rapidly as a function of beam hyper-parameter, and such behavior is hard to predict using smooth functions given the existing points in beam hyper-parameter space. It is apparent from these plots then that sub-optimal beams like the ROLSES Dipole will be difficult to reconstruct from interpolations, while the linear behavior of the Patch Antenna and the smooth behavior of the Analytical Dipole readily lend themselves to description by smooth functions in beam hyper-parameter space, such as splines. 

We employ two different interpolation methods to generate the Cryo-coefficients at beam hyper-parameter values which are not part of the original input set or grid. The first method consists of linear, cubic, and quintic spline fits for each coefficient and frequency. Several example fits using quintic splines are shown as solid curves in Figure~\ref{fig:cryo-coeff}, with each curve corresponding to the coefficient $k_l$ of the same color. For smoothly-varying functions, splines are optimal (see below for a further elucidation on this point). 

The second interpolation method we test employs Gaussian Process Regression \citep[See][for information on GPR, kernels, and optimization algorithms]{rasmussen_gaussian_2006}. Using a multi-output regressor, we fit one, independent\footnote{Note that since the beams at different frequencies and hyper-parameter values are not necessarily independent, we could instead use chains of regressors. The input parameter from one chain at one coefficient could then be used to inform the next in the chain, thus imprinting correlations from the beam patterns into the GPs. However, even for a beam at a singular hyper-parameter value, this means running a regression chain with hundreds of thousands of inputs, as one must run them for all Cryo-coefficients at all frequencies. Such correlations in chains are still an active part of GP research, so we leave such inquiries for further work and do not employ them here.}, Gaussian Process (GP) per coefficient and frequency, optimizing each GP's \textit{kernel} hyper-parameters (not to be confused with \textit{beam} hyper-parameters) using the standard L-BFGS regression algorithm \citep{liu_limited_1989}. We tested radial basis function kernels, Mat\'ern kernels with factors $\nu_\mathrm{K} = 1.5, 2.5$, and rational quadratic kernels. In all antenna cases, the Mat\'ern kernel with $\nu_\mathrm{K} = 2.5$ was adopted for the rest of this work since it performed as well as or better than the other kernels. To ensure convergence and optimal kernel hyper-parameters, we run each regression for each GP twenty times at random starting points and then select the best-fit kernel hyper-parameter values according to the highest log-marginal likelihood, which is essentially the Bayesian evidence. In our implementation we utilize the \texttt{scikit-learn} package \citep{pedregosa_scikit-learn_2018}. GPs have also found use in 21-cm foreground and systematic modeling \citep[e.g.,][]{mertens_statistical_2018,mertens_improved_2020,kern_gaussian_2020,heimersheim_flexknot_2023}. Note that we do not show the equivalent GP curves in Figure~\ref{fig:cryo-coeff}, as they overlap almost exactly with the splines on these y-axis scales.

For each antenna case we test the ability of our linear beam model to generate accurate beam patterns with respect to the corresponding CST simulation following these steps:
\begin{enumerate}
    \item Generate an input set $B_I$ of simulated or analytical beam patterns on a grid of beam hyper-parameter values with resolution $\Delta \zeta$.
    \item Decompose the input set $B_I$ into a set $K_I$ of respective Cryo-coefficients using Equation~\eqref{eqn:beam-definition} and a choice of basis $Y_{il}$.
    \item Form the test set $K_T$ of the Cryo-coefficients at one input beam hyper-parameter value $\zeta_T$, i.e., $K_T = K_I(\zeta_T)$. Remove the test set $K_T$ from the input set and denote it $K_{I-1} \equiv K_I - K_T$. 
    \item Fit splines and GP models to the truncated input set $K_{I-1}$ and denote them $M_{s}(\zeta)$ and $M_{GP}(\zeta)$, respectively. These are now the models $M_{k}$ for generating new beam patterns at any arbitrary beam hyper-parameter value $\zeta$.
    \item Generate the predicted Cryo-coefficients $K_P$ via $K_{P,GP} = M_{GP}(\zeta_T)$ or $K_{P,s} = M_{s}(\zeta_T)$. 
    \item Compare the predicted set $K_P$ to the test set $K_T$ and the corresponding beam patterns $B_P$ and $B_T$ calculated from having $Y_{il}$ acting on $K_P$ and $K_T$. 
    \item Increase beam hyper-parameter grid resolution $\Delta\zeta$ if the accuracy of the predicted coefficients or beam fall below a particular error threshold, and start again at Step 1. 
\end{enumerate}

To determine how accurately the predicted coefficients and beams from our interpolation describe the true beams and coefficients (Step 6 above), we calculate the root-mean-square (RMS) across all \texttt{Healpix} pixels of the relative error in the beam pattern, $\xi_\mathrm{RMS}$:
\begin{equation}
\begin{aligned}
    \xi_\mathrm{RMS} = \sqrt{\frac{1}{N_\mathrm{pix}}\sum_i^{N_\mathrm{pix}}\biggl| \frac{B_{T,i} - B_{P,i}}{B_{T,i}} \biggr|^2} \\ \equiv \mathrm{RMS}_{\Omega} \biggl( \biggl| \frac{B_\mathrm{CST}(\Omega) - B_\mathrm{Cryo}(\Omega)}{B_\mathrm{CST}(\Omega)} \biggr| \biggr),
    \label{eqn:relative-beam-error}
\end{aligned}
\end{equation}
where $B_T \equiv B_\mathrm{CST}$ is the test or true beam, and $B_P \equiv B_\mathrm{Cryo}$ is the predicted or interpolated beam. We also calculate the difference $\Delta T$ in antenna temperature produced when both the true and interpolated beam are weighted with a sky map created from extrapolating the Haslam $408$~MHz temperature map $T_{408}$ with a spatially constant spectral index of $-2.5$, or $T_\mathrm{sky} = T_{408}(\Omega)\nu^{-2.5}$. That is, we calculate
\begin{equation}
\begin{aligned}
    \Delta T \equiv  T_\mathrm{Ant,CST}(\nu) - T_\mathrm{Ant,Cryo}(\nu) \\ =  \frac{\sum_i^{N_\mathrm{pix }}B_{T,i}(\nu)T_{\mathrm{sky},i}}{\sum_i^{N_\mathrm{pix}} B_{T,i}(\nu)} - \frac{\sum_i^{N_\mathrm{pix}} B_{P,i}(\nu)T_{\mathrm{sky},i}}{\sum_i^{N_\mathrm{pix}} B_{P,i}(\nu)}.
    \label{eqn:ant-difference}
\end{aligned}
\end{equation}
This antenna temperature error ought to approach zero across the band as the accuracy of the Cryobeam and interpolation methods increase. 


For all antenna cases, we use a \texttt{Cryofunk} basis $Y_{il}$ generated from using a realistic observer horizon profile located at the EDGES experiment site as shown in \cite{bassett_lost_2021}. This horizon is used as it is very nearly--although not quite--flat, and so these basis functions will be very close to those for describing a $2\pi$ steradian sky. However, we note that this choice is arbitrary, and that \texttt{Cryofunk} will similarly generate a suitable basis for any desired horizon profile which can be represented in \texttt{Healpix}.

It is also useful to understand how interpolation in Cryo-coefficient space compares to interpolation of the linear gains directly in the raw CST beam data format in spherical coordinates $(\theta, \phi)$, which we shall refer to as spherical coordinate pixel interpolation. If the gain patterns at a particular coordinate $(\theta,\phi)$ vary smoothly with beam hyper-parameter value $\zeta$, we would expect spherical coordinate pixel interpolation to produce beams at least as accurate as when interpolating in Cryo-coefficient space. Therefore, direct spherical coordinate pixel interpolation provides another measure of the smoothness of the beam. Thus, we interpolate each antenna case's linear gain patterns in spherical coordinate pixels using the splines noted above for all spherical coordinate pixels above the horizon, and then calculate the aforementioned error metrics.

In addition, we have also tested interpolating the beams in \texttt{Healpix} pixel-space after converting the raw spherical coordinate pixel beams to \texttt{Healpix} pixel-space. This \texttt{Healpix} pixel-space beam interpolation yields relative beam errors nearly identical to the spherical coordinate pixel beam interpolation, i.e., the relative difference of the beam-weighted spectrum of a spatially uniform $10^4$~K foreground between the spherical coordinate pixel beam and \texttt{Healpix} pixel beam is on the order of $10^{-12}$, for beams with roughly the same resolution ($1^{\circ}\times1^{\circ}$ angular resolution $\sim N_\mathrm{side} = 64$), and when we degrade the \texttt{Healpix} pixel-based beams to $N_\mathrm{side} = 32$. Therefore, only the spherical coordinate pixel beam interpolations are used in this study.

From these initial tests of Cryofunk and spherical coordinate pixel interpolations, there are several observations and comparisons to be made:
\begin{enumerate}[(i)]
    \item The beam pattern may vary more smoothly in one space versus another (Cryofunk versus spherical coordinate pixels), potentially making interpolation more accurate. For example, although both the spherical coordinate pixel gain values (Figure~\ref{fig:polar-plots}) and Cryo-coefficients (Figure~\ref{fig:cryo-coeff}) of the ROLSES Dipole exhibit non-smooth variation as a function of $\epsilon$, interpolation in the Cryo-coefficient space is more accurate at various grid resolutions (see Section~\ref{sec:results-beam-em} below). We acknowledge that it is difficult to determine \textit{a priori} which interpolation scheme will work better (without e.g. plotting hyper-parameter dependencies), and not all antenna beams will necessarily be interpolated better in Cryo-coefficient space. As shown in Section~\ref{sec:results-beam-em}, however, interpolation of Cryo-coefficients for the antenna cases studied here is always more accurate than or comparable to spherical coordinate pixel interpolation.
    \item The \texttt{Cryofunk} basis $Y_{il}$ gives one the ability to examine how particular basis shapes, or equivalently basis vectors (columns of $Y_{il}$), describe a beam pattern, allowing one to determine which basis shapes dominate its structure (see Section~\ref{sec:discussion} for a discussion of how many modes are needed for a given level of accuracy). In direct spherical coordinate pixel interpolation there is obviously no such basis, as each coordinate is being interpolated separately; however, degrading the resolution of the beam in spherical coordinate pixels would be analogous to truncating the number \texttt{Cryofunk} basis vectors.
    \item Model evaluation time is about $\sim 1.5$ times faster using Cryo-coefficients ($\sim4.7~$ms) than spherical coordinate pixels ($\sim7~$ms) for beam interpolation. When Cryofunk is combined with a typical foreground model, as in the next section, model evaluation time can improve up to an order of magnitude since the Cryo-basis and foreground maps can be combined beforehand.
\end{enumerate}
The Cryo-coefficient arrays used for interpolation are smaller than the spherical coordinate pixel arrays ($1^{\circ}$ and $2^{\circ}$ spherical coordinate angular resolutions correspond to $\sim$~32,400 and 8,100 elements above the horizon to interpolate, compared to $\sim$ 25,000 and 6,000 elements for corresponding \texttt{Healpix} maps with $N_\mathrm{side} =$ 64 and 32, respectively). Secondly, as noted above, the \texttt{Cryofunk} basis $Y_{il}$ can be combined with the foreground model as in Equation~\eqref{eqn:bwf-cryo-model}, allowing one to drastically reduce model evaluation time by precomputing the Cryo-chromaticity functions (see the next section for details). In contrast, in the direct spherical coordinate pixel case the entire beam pattern must be interpolated and formed at every model evaluation step \textit{and then} weighted with the foreground and horizon models.

Lastly, we note that both spline and GP fits to the test sets $K_{I-1}$ in Cryo-coefficient space and direct spline fits to spherical coordinate pixel space can be tested for accuracy by calculating the coefficient of determination, also known as the \textit{score} or $R^2$ value, in this work given by
\begin{equation}
    R^2 = 1 - \frac{\sum_i \left[k_l(\zeta_i) - M_k(\zeta_i)\right]^2}{\sum_i \left[k_l(\zeta_i) - \Bar{k}_l(\zeta_i)\right]^2},
\end{equation}
where the numerator is essentially the sum of the squared residuals for a given interpolation model $M_{s}$ or $M_{GP}$, and the denominator is the sum of squared residuals between the input coefficients and their mean value $\Bar{k}_l$. A score of unity denotes the case when the modeled values fit the input values exactly. The splines used in this work generated $R^2$ values of unity for cubic and quintic orders, while the GPs tended to produce $R^2$ levels below one, around $0.8$ or greater for the highest resolution grids and Mat\'ern kernels. The score then is the measure that we are referring to when stating that splines are optimal for describing smooth beam data presented in this work. 

\section{Fitting Radio Spectrometer Data}
\label{sec:methods-fitting}

\subsection{Beam-Weighted Foreground Model}

Of the advantages offered by using a linear beam model, the first is the increased speed of model evaluations, and the second, largely abetted by the first, is the ease with which it can be combined with models for other systematics. This is especially useful in the context of global 21-cm cosmology. A beam-weighted foreground model, used for instance in both \cite{anstey_general_2021} and \cite{hibbard_fitting_2023}, is given by
\begin{equation}
    \mathcal{M}_\mathrm{FG}(\nu) = \sum_j^{N_r} K_j(\nu) A_j \left( \frac{\nu}{\nu_{0}}\right)^{\beta_j + \gamma_j \ln(\nu / \nu_0)} + T_\mathrm{CMB},
\end{equation}
with 
\begin{equation}
    \begin{aligned}
        K_j(\nu) =&~ \frac{1}{4\pi} \int_0^{4\pi} B(\Omega, \nu) H(\Omega) M_j(\Omega) \\ & \times \biggl\{\frac{1}{2N_n\delta t} \sum^{N_n}_{i=1} \int^{t_i + \delta t}_{t_i - \delta t} \bigl[T_{0}(\Omega, t) - T_\mathrm{CMB}\bigr] \diff t\biggr\}\diff\Omega.
    \end{aligned}
    \label{eqn:chromaticity-functions}
\end{equation}
where $B$ and $H$ are once again the beam pattern and horizon profile respectively, $M_j$ is the mask of the foreground map for sky region $j$, $T_0$ is the base brightness temperature map used in extrapolation, typically the 408 MHz map \citep{haslam_408-mhz_1982,remazeilles_improved_2015}, $T_\mathrm{CMB} = 2.725$~K is the CMB temperature at the present time, $\delta_t$ represents the integration time, and $N_n$ the number of spectra sampled and averaged together within a local sidereal time (LST) bin $n$. Lastly, the intrinsic foreground parameters are given by the temperature magnitude in each region $A_j$, the spectral index $\beta_j$, and the spectral curvature $\gamma_j$.

The term in curly brackets in Equation~\eqref{eqn:chromaticity-functions} is only dependent upon frequency and space after the time integral is performed. This term can be precomputed beforehand, and we label it $\mathcal{T}(\Omega)$. Working again in \texttt{Healpix}, Equation~\eqref{eqn:chromaticity-functions} becomes
\begin{equation}
    \mathcal{M}_\mathrm{FG}(\nu) = \sum_j^{N_r} \biggl[\sum_i^{N_{pix}} B_i(\nu, \boldsymbol{\zeta}) H_i M_{ji} \mathcal{T}_i \biggr] F_j(\nu) + T_\mathrm{CMB},
    \label{eqn:FGmodel_notreduced}
\end{equation}
where we have defined $F_j(\nu) \equiv A_j \left( \frac{\nu}{\nu_{0}}\right)^{\beta_j + \gamma_j \ln(\nu / \nu_0)}$.

Using our linear beam model in Equation~\eqref{eqn:base-cryo-beam}, Equation~\eqref{eqn:FGmodel_notreduced} can be rewritten as
\begin{equation}
\begin{aligned}
\mathcal{M}_\mathrm{FG}(\nu) &= \sum_j^{N_r} \sum_l^{N_b} \left( \sum_i^{N_H} Y_{il} M_{ji} \mathcal{T}_i \right) k_l(\nu, \boldsymbol{\zeta}) F_j(\nu) + T_\mathrm{CMB}.
\end{aligned}
\label{eq:}
\end{equation}
In such a format, we can graciously precompute the Cryo-chromaticity functions in the \texttt{Cryofunk} basis, within the parenthesis, instead of computing them at each model evaluation. They are now a matrix which depends upon the spectral region $j$ and the \texttt{Cryofunk} basis coefficient $l$. Writing these as $\mathcal{C}_{jl}$, our final beam-weighted foreground equation is concisely,
\begin{equation}
    \mathcal{M}_\mathrm{FG}(\nu) = \sum_j^{N_r} \sum_l^{N_b} \mathcal{C}_{jl} k_l(\nu, \boldsymbol{\zeta}) F_j(\nu) + T_\mathrm{CMB}.
    \label{eqn:bwf-cryo-model}
\end{equation}

\subsection{Likelihood and Nonlinear Sampling \\ with \texttt{MEDEA}}

In 21-cm cosmology, the beam is a key systematic hampering efforts to extract the weak cosmological 21-cm emission signal from the much brighter synchrotron foreground. It is thus of singular importance to determine if constraints on the beam hyper-parameters $\boldsymbol{\zeta}$ (and thus on the beam pattern itself) and the galactic foreground parameters can be extracted from spectra synthesized from realistic galactic foregrounds weighted by Cryobeams.

We utilize as the intrinsic foreground (for 1 LST bin) that of \cite{hibbard_fitting_2023} generated at the frequencies relevant for each of our antenna cases to compute mock beam-weighted foreground spectra $T_\mathrm{mock}(\nu)$  using each respective beam $B_P(\zeta_T)$ \citep[for a comparable example, see Figure~1 of ][]{hibbard_fitting_2023}. Given this mock beam-weighted foreground and a relevant observational noise level, we form the usual log likelihood assuming Gaussian distributed noise as
\begin{equation}
    \ln{\mathcal{L}} \propto -\frac{1}{2}(\boldsymbol{T}_\mathrm{mock} - \boldsymbol{\mathcal{M}}_\mathrm{FG})^T \boldsymbol{C}^{-1} (\boldsymbol{T}_\mathrm{mock} - \boldsymbol{\mathcal{M}}_\mathrm{FG})\,,
    \label{eqn:likelihood}
\end{equation}
where $\boldsymbol{\mathcal{M}}_\mathrm{FG}$ is given by Equation~\eqref{eqn:bwf-cryo-model}. The bold-faced symbols are vectors with dimensionality equal to the number of frequency channels, except for $\boldsymbol{C}^{-1}$ which is the noise covariance matrix with $\sigma_n^2$ along the diagonal. The parameters to be fit in this model are the galactic foreground parameters $A_j$, $\beta_j$, and $\gamma_j$ for each spectral region $j$, and the beam hyper-parameter $\zeta$.

The total number of parameters in each model evaluation is thus $N_r + 1$, where $N_r$ is the number of spectral regions. Note that our beam-weighted foreground model contains only the subset of Cryo-coefficients $K_{I-1}$, so that our model for the interpolation of Cryo-coefficients is given by $M_s(\zeta)$, as noted above. Thus, we are not only testing how well we can extract the true beam hyper-parameter value $\zeta_T$ from beam-weighted spectra for Cryo-coefficients not included in the model, but are also exploring the level of covariance between the beam hyper-parameters and the intrinsic foreground parameters.

In the simplest case, the beam-weighted foreground model (Equation \ref{eqn:bwf-cryo-model}) is assumed to be the correct model to fit the mock spectra. The latter then contains only thermal noise fluctuations, and can be described by a diagonal covariance matrix $\boldsymbol{C}^{-1}$ with $\sigma^2_n$ denoting the thermal noise level. However, measured spectra from an experiment will likely differ at some level from the beam-weighted foreground model assumed here---i.e., the model will exhibit bias. This is unavoidable when fitting experimental data, and therefore models ought to be able to absorb this bias either by e.g. increasing the model complexity, the number of model parameters, or the model parameter ranges. Model biases and three ways in which \texttt{MEDEA} can be used in future work to potentially account for model bias when fitting real radio spectrometer data are discussed in Section~\ref{sec:discussion} .

As an additional consistency test beyond the thermal noise fits, a \textit{systematic bias} fit is performed to the Analytical Dipole antenna, where we include a systematic bias in the mock spectra that is not included in the beam-weighted foreground model. We introduce this bias into the mock by computing a perturbed Analytical Dipole using Equation \ref{eqn:dipolegroundplane} but with $\theta \rightarrow \theta + \delta\theta$, where $\delta \theta = 1^{\circ}$ represents a systematic, a one-degree offset in the beam pattern. Such a systematic could be plausibly caused by, for example, unaccounted-for ionospheric effects \citep{shen_quantifying_2021}. We denote this example fit to the perturbed Analytical Dipole the systematic bias case, and note that the covariance matrix is the same here as in the thermal noise case. This case is intended to be illustrative of how inadequate models, without proper optimization, flexibility, or parameters, are expected to produce biased fits. We later discuss in Section \ref{sec:discussion} how such biases can be accounted for, especially in the context of \texttt{MEDEA}.

We use the nested sampling algorithm \texttt{Polychord} \citep{handley_polychord_2015} with live points $n_\mathrm{live} = 100 \times (N_r + 1)$ to produce posterior constraints on both the galactic foreground parameters and the beam hyper-parameters. We run this nonlinear sampling procedure for various hyper-parameter grid resolutions and number of spectral regions. For simplicity and as an initial test, the only galactic foreground parameters we vary in this work are the spectral indices $\beta_j$.

\section{RESULTS: Beam Emulation}
\label{sec:results-beam-em}
This Section~summarizes the main results of using the linear beam model \texttt{MEDEA} for interpolation and fitting beam-weighted foreground spectra.

\begin{figure*}
    \centering
    \includegraphics[width=0.8\textwidth,height=7cm]{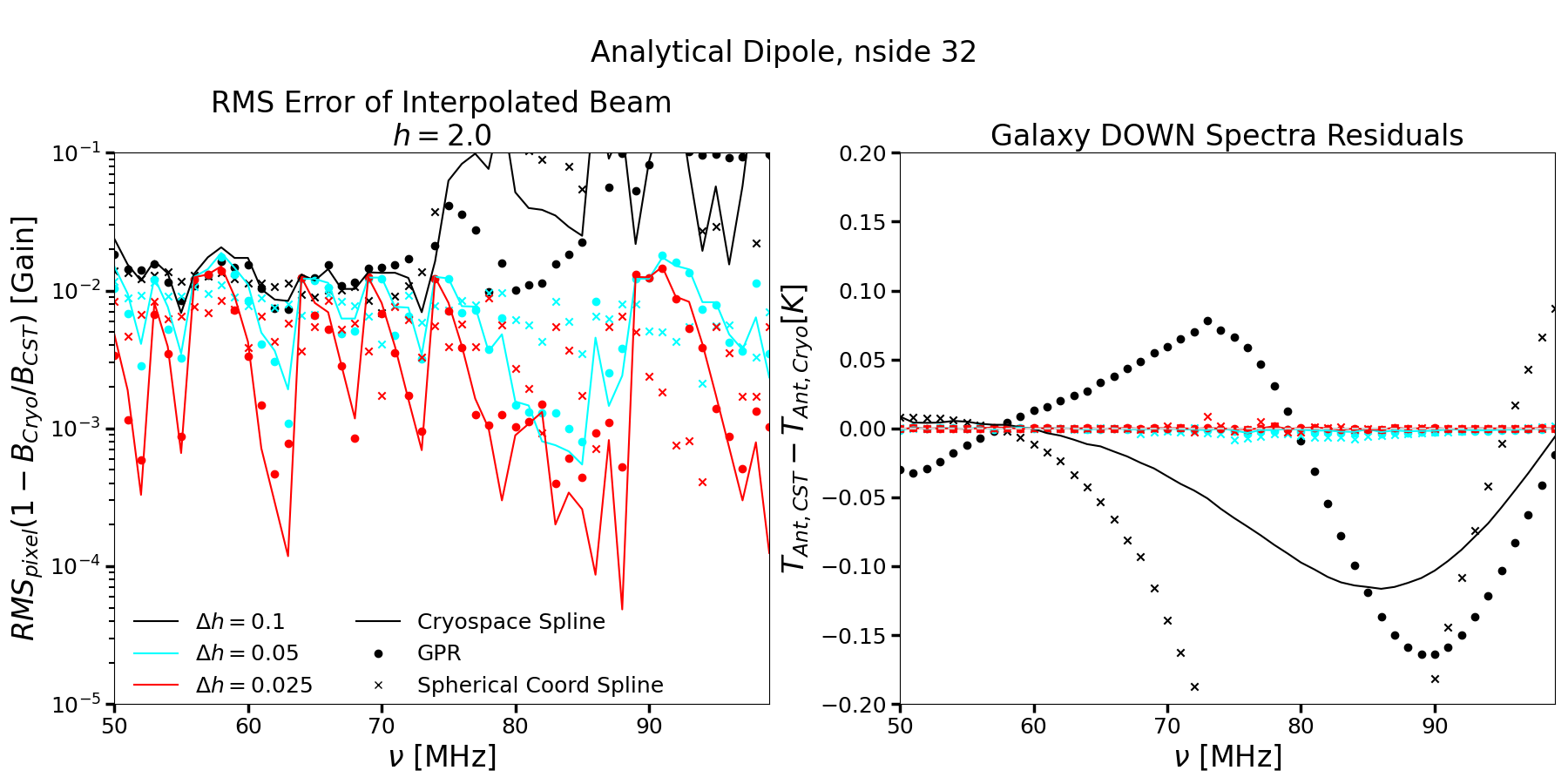}
    \includegraphics[width=0.8\textwidth,height=7cm]{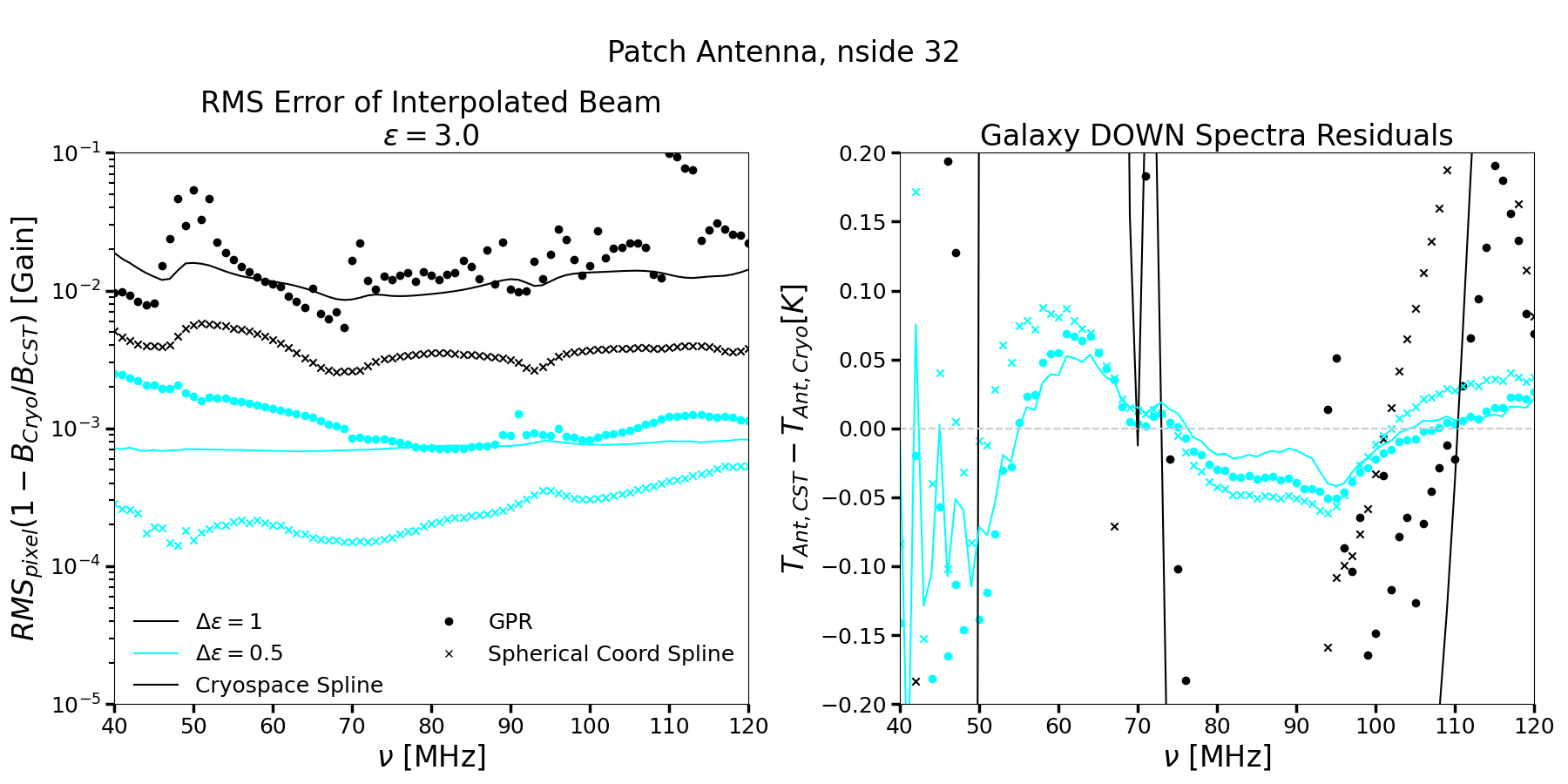}
    \includegraphics[width=0.8\textwidth,height=7cm]{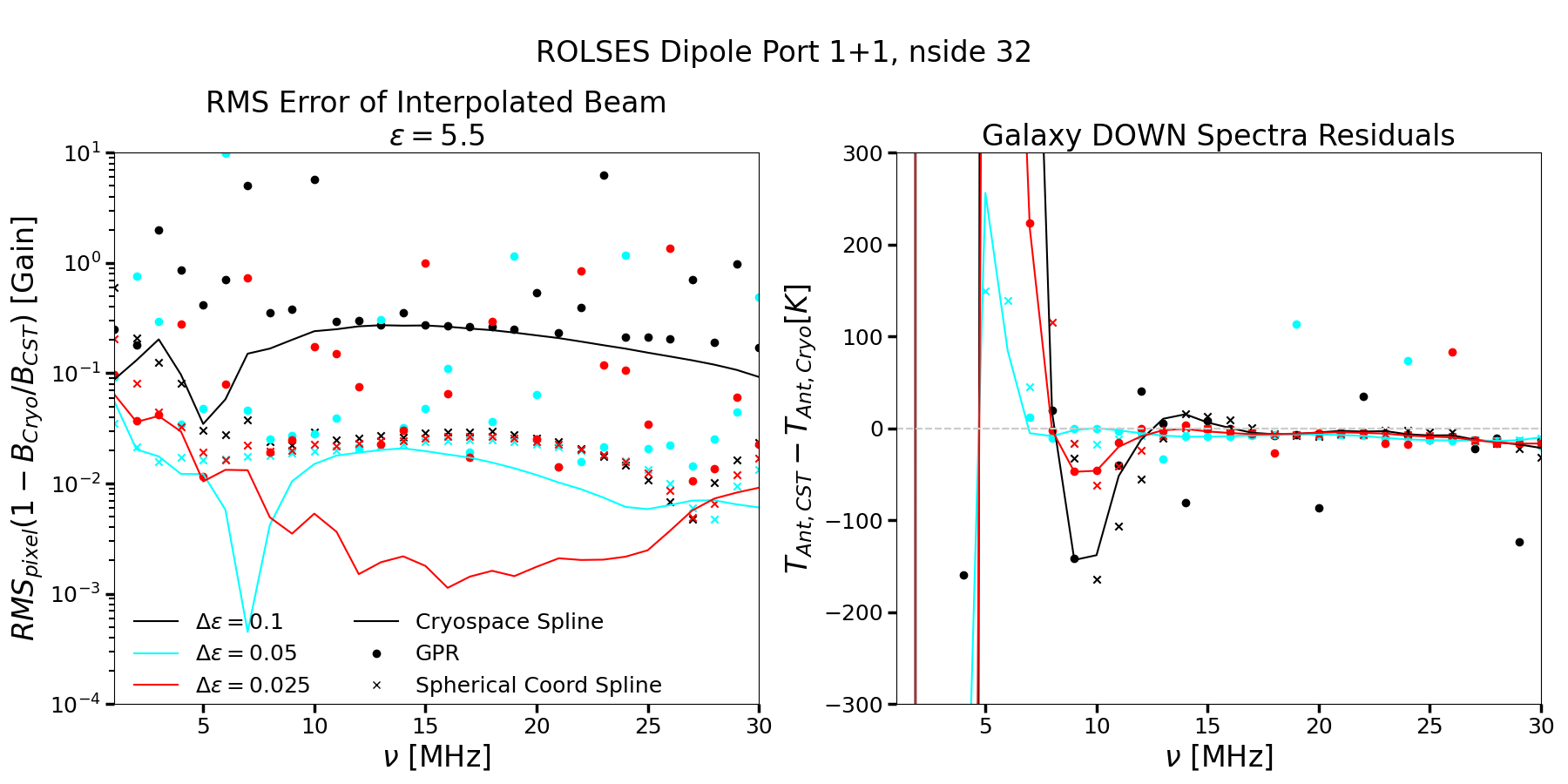}
    \caption{Interpolation results from using Cryo-coefficient splines (solid lines), Gaussian Processes (circles), and splines directly to the gains in spherical coordinate pixels (crosses) for all antenna cases. Each color denotes a different hyper-parameter grid resolution, from low (black) to medium (cyan) to high (red). The left column shows the relative beam error (Equation~\ref{eqn:relative-beam-error}) between the interpolated Cryobeam from \texttt{MEDEA} and the true CST beam. Note that the RMS over all beam pixels is computed for the relative error. The right column shows the difference in antenna temperatures between the interpolated Cryobeam and the true CST beam when both weight the same intrinsic foreground map (Equation~\ref{eqn:ant-difference}). The light gray, dashed line indicates the zero point for each antenna case.}
    \label{fig:interp-results}
\end{figure*}

Figure~\ref{fig:interp-results} summarizes the main results for all three antenna cases and each method of interpolation for one test hyper-parameter example; nonetheless, our results are qualitatively unchanged when different test hyper-parameters are used for each antenna case. Plotted in the left column of each panel is the RMS (over beam pattern pixels) of the relative beam error calculated using Equation~\eqref{eqn:relative-beam-error}, while the right column depicts the difference between the CST and Cryobeam antenna temperatures when both are weighted with the Haslam galaxy, as obtained using Equation~\eqref{eqn:ant-difference}. In each panel, the different colors correspond to increasing grid resolutions for each beam hyper-parameter, from black (lowest), to cyan (middle), to red (highest). The different curves correspond to different methods of interpolation:
\begin{itemize}
    \item Solid curves - Spline interpolations in Cryo-coefficient space, 
    \item Filled circles - GP interpolations in Cryo-coefficient space, 
    \item Crosses - Spline interpolation directly to the gain in spherical coordinate pixels. 
    \end{itemize}    
As stated, the left column in each panel shows the RMS over pixels of the relative beam error, and so smaller values correspond to more accurate both interpolations and respective Cryobeams. The right column depicts the difference between the two antenna temperatures from the resultant beams and thus ought to approach zero for more accurate interpolations produced by \texttt{MEDEA}.

It is clear that increasing the grid resolution increases the accuracy of \texttt{MEDEA} and the Cryobeam it outputs, as expected. For the Analytical Dipole (top panel), Cryobeams with grid resolutions of $\Delta h = 0.05$~m and $\Delta h = 0.025$~m exhibit errors of order $\sim 10^{-2}~(\lesssim-20~\mathrm{dB})$ and $\sim 10^{-3}$ respectively (left graph). These relative errors correspond to antenna temperature differences (right graph) which are below the noise levels often cited in Cosmic Dawn experiments, $\leq 20$~mK. This beam interpolation accuracy is achieved with 21 and 41 beams, respectively. Moreover, the Patch Antenna (middle panel) relative beam errors are of order $10^{-2}$ and $10^{-3}$ for grid resolutions of $\Delta \epsilon = 1$ and $\Delta \epsilon = 0.5$, respectively, although it has less than half the number of beam patterns in the input set compared to the Analytical Dipole. The Patch Antenna temperature differences (right plot) are around $\sim 50$~mK at these resolutions. We note again that the number of Patch Antenna simulations run was determined by the computational resources dedicated to this work, but that running and utilizing a higher number would certainly increase the accuracy of the beams produced by \texttt{MEDEA}. The ROLSES Dipole achieves a similar level of relative beam error using splines, yet exhibits much larger antenna temperature residuals ($\sim 100$~K in the right column of the bottom panel) owing to the much greater galaxy temperatures present at the lower frequencies ($1-30$~MHz) relevant for Dark Ages (DA) experiments. Therefore, we find that a beam precision of much less than -30~dB is needed for DA experiments.

Comparing the two Cryo-coefficient space interpolation methods, we see that the GP tends to perform as well as the splines for both the Analytical Dipole and the Patch Antenna, but suffers when applied to the ROLSES Dipole. Given the erratic and non-smooth coefficients present in the latter's Cryo-coefficient space decomposition (see Figure~\ref{fig:cryo-coeff}), this is perhaps not surprising, as GPs with Mat\'ern kernels perform better for fitting smoothly varying functions \citep{wang_intuitive_2020}. Furthermore, the oscillatory nature of the ROLSES Dipole beam pattern (bottom panel, Figure~\ref{fig:polar-plots}) explains the comparative failure of the direct spherical coordinate pixel splines (crosses) to produce relative beam errors at the same level as the Cryo-coefficient space splines. The Cryo-coefficients are simply smoother functions of the beam hyper-parameter than the gains in spherical coordinate pixels themselves, so the predictive power of the smooth splines is increased in Cryo-coefficient space. We note, however, that the direct gain spherical coordinate pixel splines perform at roughly the same level (to within a factor of three or better) as the Cryo-coefficient splines for the Analytical Dipole and Patch Antenna, owing to their smooth beam pattern structures. Note again that these spherical coordinate pixel interpolations are for one degree resolution beam patterns. This shows that the Cryo-coefficient splines produce relative beam errors at the same, or lower, level as direct gain spherical coordinate pixel interpolations. Furthermore, this demonstrates the efficacy of the performance of \texttt{MEDEA}, as anticipated in Section~\ref{sec:methods-beam-interp}.

\begin{figure*}
    \centering
    \includegraphics[width=0.96\textwidth]{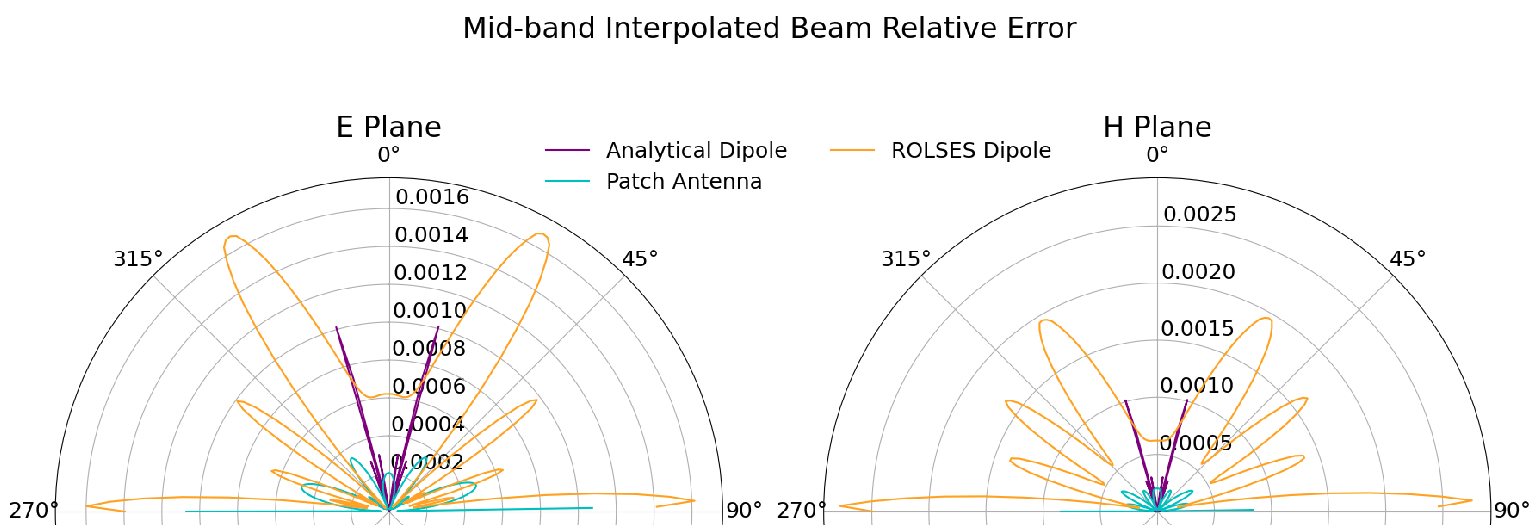}
    \caption{Polar plots showing the $E$- and $H$-plane cuts for the relative beam errors (Equation~\ref{eqn:relative-beam-error} without calculating the RMS over pixels) for all three antennas at mid-band frequencies. Note that these are the beams generated at the missing hyper-parameter value $\zeta_T$ for each case.}
    \label{fig:polar-plot-interp-beam}
\end{figure*}

In Figure~\ref{fig:polar-plot-interp-beam}, we show an example polar plot projection of the relative beam errors using Equation~\eqref{eqn:relative-beam-error} without calculating the RMS over pixels, when using \texttt{MEDEA} to generate the missing beam in the set at hyper-parameter value $\zeta_T$. These polar plots, shown at the mid-band frequency for each antenna case, depict the (relative) residuals between the CST beam and the \texttt{MEDEA} beam, highlighting the leftover structures not captured by our model. These structures are very small, below $0.2\%$ error for all cases, in accordance with the RMS values plotted in Figure~\ref{fig:interp-results}. The residual structures are dominated by the Cryo-coefficients $k_l$ with the highest interpolation error, and thus their corresponding vectors in the basis $Y_{il}$. As a result, these structures are not necessarily noise-like.

\section{RESULTS: Fitting Radio Spectrometer Data}
\label{sec:results-fitting}

The main results from fitting our likelihood in Equation~\eqref{eqn:likelihood} using our beam-weighted foreground model and beam hyper-parameters are summarized in Table \ref{tab:spec-fit-stats}. For the Analytical Dipole, we find that $N_r < 5$ spectral regions are necessary to fit the mock foreground spectra down to the noise level for 1~LST bin for the thermal noise case, and we find that the middle resolution beam hyper-parameter grids are sufficient for interpolation. The numerically simulated beams, in contrast, require many more spectral regions ($N_r > 10$), although the middle resolution parameter grids are also adequate to fit our models down to the noise level. With higher noise levels, however, an even sparser resolution is acceptable, as shown in the right columns of Figure~\ref{fig:interp-results}.~Note for example that for noise levels corresponding to 1~K, the lowest resolution grids are suitable. This precision level is more than sufficient for antenna design studies. The last column of Table \ref{tab:spec-fit-stats} shows the marginalized one-dimensional posterior means and standard deviations of the beam hyper-parameters from the nonlinear fits to the spectra.
\begin{table*}[htb!]
    \centering
    \begin{tabular}{c|c|c|c|c|c|c|c|c}
    \hline
    Antenna & No. of & Grid & Input & Log & Input & Posterior & Posterior & Bias \\
    Case & Regions & Resolution & Noise & Evidence & Hyper-parameter & Mean & Error & \\
    & $N_r$ & $\Delta \zeta$ & $\sigma_n$ & $\ln{Z}$ & $\zeta$ & $\mu_{\zeta}$ & $\sigma_{\zeta}$ & $|\zeta - \mu_{\zeta}|$ \\
    \hline
    Analytical Dipole & 4 & $\Delta h = 0.05$ & 20 mK & -70 & 2 & 1.99 & 0.001 & 0.01 \\
    Patch Antenna & 11 & $\Delta \epsilon = 0.50$ & 50 mK & -108 & 3 & 2.99 & 0.01 & 0.01 \\
    ROLSES Dipole & 13 & $\Delta \epsilon = 0.05$ & $\sigma_{rad}$ & -878 & 5.5 & 5.49 & 0.001 & 0.01 \\
    \hline
    Pert. Analytical Dipole & 4 & $\Delta h = 0.05$ & 20 mK & -3051 & 2 & 2.054 & 0.0001 & 0.054 \\
    \hline
    \end{tabular}
    \caption{This table shows the results of using \texttt{MEDEA} to fit mock beam-weighted foreground spectra. $N_r$ refers to the number of spectral regions and hence spectral indices fit alongside the beam hyper-parameter. The Grid Resolution describes the beam hyper-parameter resolution used in \texttt{MEDEA}. The input noise level, $\sigma_n$, for the ROLSES Dipole, denoted by $\sigma_{rad}$, corresponds to radiometer noise for 1 MHz channel spacing and $8$~days of integration time. The fifth column denotes the log evidence produced by the nonlinear fit, the sixth shows the input beam hyper-parameter, while the beam hyper-parameter means and standard deviations ($68 \%$ confidence levels) are reported in the seventh and eighth columns and are generated from the marginalized posterior samples. The final column shows the bias, or absolute difference between the input beam hyper-parameter and posterior mean. The last row shows the results from using \texttt{MEDEA} to fit the perturbed Analytical Dipole, denoted the systematic bias case. The systematic bias is introduced into the mock beam-weighted foreground spectrum in the form of an offset of all $\theta$ coordinates producing the beam by one degree. Such an offset might represent, for example, unaccounted-for effects of the ionosphere.}
    \label{tab:spec-fit-stats}
\end{table*}

\begin{figure}
    \centering
    \includegraphics[width=0.46\textwidth]{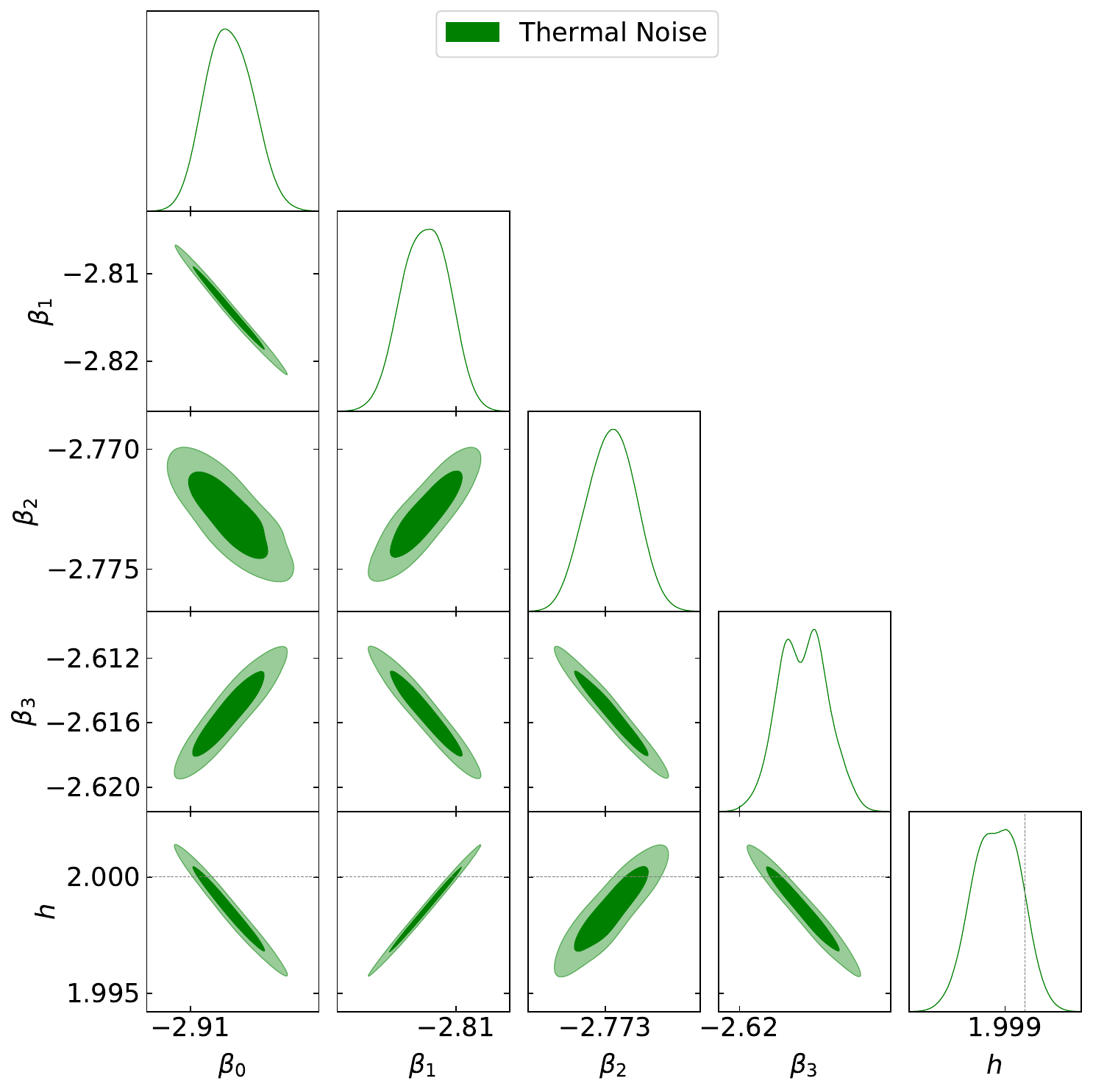}
    \caption{Triangle plot showing the galactic foreground spectral index parameters ($\beta_i)$ and beam hyper-parameter $h$ for the Analytical Dipole fit. The mock beam-weighted foreground spectra were made using a beam hyper-parameter of $h = 2.0$~m. The contours here correspond to the thermal noise case, where the mock radio spectrometer data is well-described by the beam-weighted foreground model of Equation \ref{eqn:bwf-cryo-model}. The black dotted line indicates the input beam hyper-parameter value used to generate the mock radio spectrometer data.}
    \label{fig:posterior-analytical-dipole}
\end{figure}

In Figure~\ref{fig:posterior-analytical-dipole}, we show an example triangle plot for the Analytical Dipole fit with $N_r = 4$ spectral regions and one beam hyper-parameter, $h$. The parameter contours shown in Figure~\ref{fig:posterior-analytical-dipole} are created from the posterior samples generated by \texttt{Polychord} during the nonlinear fitting procedure. The Analytical Dipole beam hyper-parameter for the thermal noise case, in particular, is well-constrained at a $68\%$ confidence level to $< 0.002$~m, despite strong degeneracies with the spectral indices $\beta_j$, as seen in the final row of the top panel of Figure~\ref{fig:posterior-analytical-dipole}. We note again that the model in this fit does not contain the beam at $\zeta_T = h = 2.0$~m used to create the mock  beam-weighted foreground. However, we can generate such a beam (and therefore deduce the underlying hyper-parameter value) by using our linear beam model, with which we can continuously fit the beam-weighted foreground spectra. It is remarkable that this beam pattern can be reconstructed from spectra without \textit{a priori} spatial information. For the other antenna cases, we can also constrain the beam hyper-parameters at their respective noise levels.

The main results from the systematic bias case for the perturbed Analytical Dipole are shown in the last row of Table \ref{tab:spec-fit-stats}. We remind the reader that the latter case is generated by introducing a one degree offset in the beam pattern in all $\theta$ coordinates of the Analytical Dipole, representing a systematic bias due to an unaccounted-for effect, such as e.g. the ionosphere. As apparent from the posterior mean and error column, there is a large bias in the extracted beam hyper-parameter $h = 2.054$, as expected. Furthermore, the $68\%$ confidence level for the posterior beam hyper-parameter in the systematic bias fit is spuriously small at $0.0001$, especially given the level of bias of the mean. The posterior contours of the spectral indices are similarly small, and they cannot even be straightforwardly seen if they were to be plotted at the parameter ranges shown in Figure~\ref{fig:posterior-analytical-dipole}. Again we emphasize that such results are expected given the unmitigated bias introduced. This fit could be optimized in terms of number of spectral regions, parameters, Bayesian evidence, etc. This case thus illustrates the evident need for flexible models that can absorb systematic bias. See Section~\ref{sec:discussion} for a discussion on ways to extend \texttt{MEDEA} to fit real radio spectrometer data and potentially account for such biases.

Lastly, in Appendix \ref{app-patch-ant-triangle-plot}, we show Patch Antenna posteriors from fitting the beam-weighted foreground spectrum using \texttt{MEDEA} (Figure~\ref{fig:patch-rectangle}). Note that we only show the contours depicting the covariances between the beam hyper-parameter $\epsilon$ and the foreground spectral indices $\beta_j$. These contours are largely ellipsoidal, indicating small degeneracies between the regolith dielectric constant and the spectral indices, for this antenna case.

\section{Further Extensions}
\label{sec:discussion}
When fitting real radio spectrometer data from experiments, it is expected that the beam simulations, and thus \texttt{MEDEA} models, will be different from the true beam used for the observation. Hence, we would expect the extracted parameter contours, such as for the beam hyper-parameter or galactic spectral indices, to be biased or exhibit spuriously small error levels. To account for this kind of systematic bias, one typically increases the flexibility of the model by increasing the number of parameters that are allowed to vary or by widening the parameter prior ranges. Although beyond the scope of this study and left for future work, this Section~identifies some of the potential extensions to \texttt{MEDEA} to allow it to account for these types of systematics and biases.

There are at least three separate ways to increase the number of parameters in, or model flexibility exhibited by, \texttt{MEDEA}: 1) allow the Cryo-coefficients themselves to vary within well-defined prior ranges and preserve the relative correlations which are generated by the decomposition into Cryo-space; 2) use \texttt{MEDEA} to generate a large number of beam simulations, weight them with foreground models, and then construct linear modes from the resultant training set of beam-weighted foreground spectra using SVD; 3) generate more input beams by varying other parameters of the beam simulation (such as generating more beams with various soil $\tan \delta$ values).  In each case, these additional parameters can be potentially utilized to account for the bias inherent when fitting models to real experimental data.

\subsection{Extensions of the Linear Beam Model \texttt{MEDEA}}
Firstly, as the model consists of coefficients which describe the amount of power for particular shapes in a beam pattern, these coefficients could be fit as parameters in a likelihood, rather than interpolated between pre-computed values. Priors could be set on each coefficient $k_l$ by forming a distribution from a set $K_I$ of Cryo-coefficients obtained from CEM simulations. However, the obvious drawback to this method is the extremely high number of Cryo-coefficients which may be necessary to include in a fit down to the $-20$~dB level or less. Furthermore, the coefficients will exhibit strong degeneracies with the parameters used in some galactic foreground models, such as those presented in \citep{anstey_use_2023,hibbard_fitting_2023}. The amplitude parameters multiplying the base temperature map will be especially degenerate with the beam coefficients. We leave an exploration of these parameter degeneracies, however, to future work.

\begin{figure}
    \centering
    \includegraphics[width=0.46\textwidth]{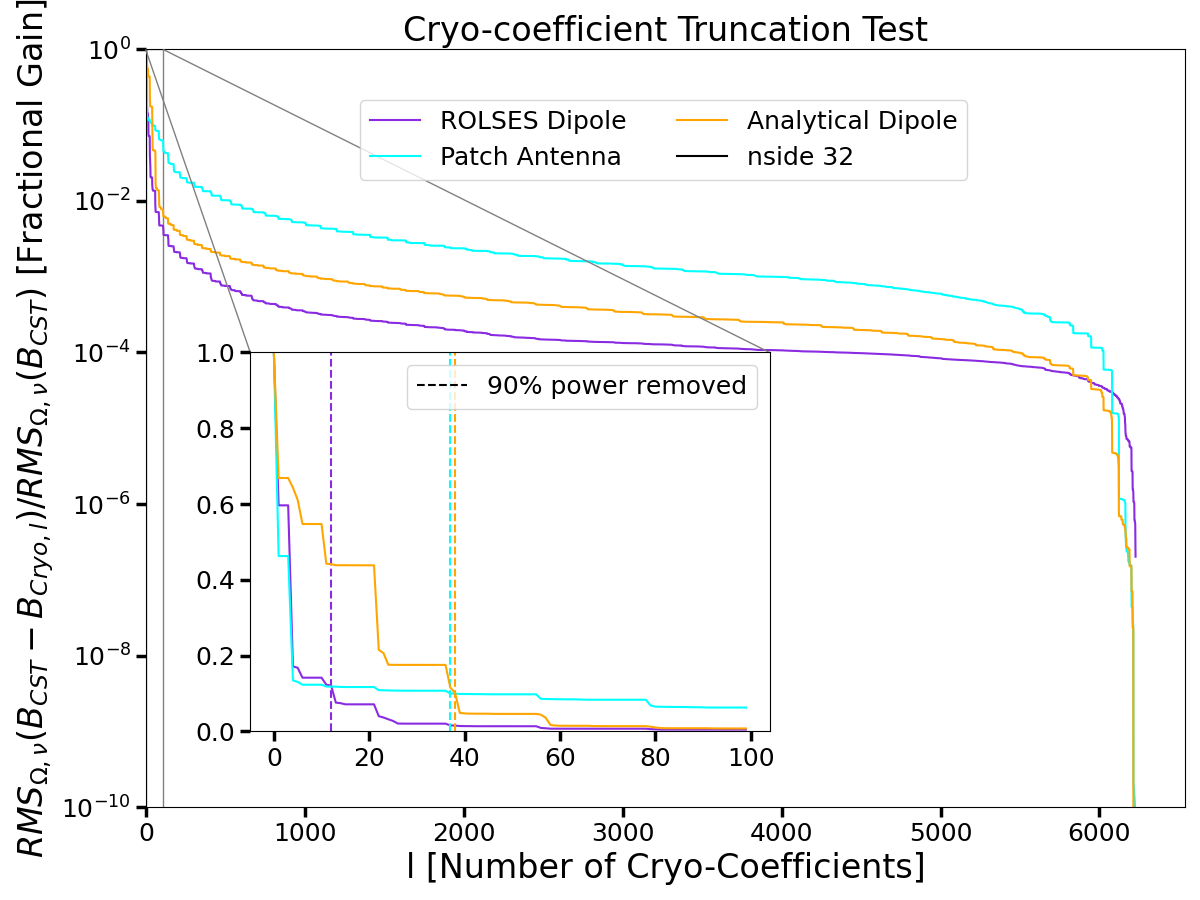}
    \caption{Results from truncating the \texttt{Cryofunk} basis to the number of terms shown on the x-axis for each antenna case. Each color denotes a different antenna at a \texttt{Healpy} resolution of $N_\mathrm{side} = 32$: the ROLSES Dipole (purple), the Patch Antenna (cyan), and the Analytical Dipole (gold). The y-axis represents the amount of gain removed by each Cryo-mode, normalized to the RMS of the full gain of the CST beam. Therefore, the y-axis is unity and the fractional gain removed is zero when no Cryo-modes are included in the Cryobeam; in contrast, the fractional gain approaches zero when all modes are included in the Cryobeam, as the Cryobasis is a complete basis. As can be seen from the inset, 20-40 modes removes approximately $90 \%$ of the gain for all cases, and most of the gain is removed by $\sim 100$ modes. While the remaining modes above $l=100$ remove several orders of magnitude of power, all of them are well below $<10^{-5}$, so they contribute comparatively little to the overall beam pattern.}
    \label{fig:trunc-test}
\end{figure}

In Figure~\ref{fig:trunc-test}, we plot the fractional gain removed from the fiducial CST beam $B_\mathrm{CST}$ by $B_\mathrm{Cryo,l}$ as a function of the number of Cryo-coefficients $l$ used to generate $B_\mathrm{Cryo,l}$. This fractional gain is normalized to be unity when no Cryo-coefficients are included in the Cryobeam and approaches zero when all coefficients (or modes) are included in the Cryobeam, as the Cryobasis is a complete basis. As expected, the first $\sim 100$ modes for each antenna case are the most important, as they remove the greatest relative magnitude of gain from the fiducial CST beam. Indeed, between $20-40$ modes are required to remove $90 \%$ of the gain for the antenna cases in this work, as shown in the inset of Figure~\ref{fig:trunc-test}.

In contrast, the highest-order Cryo-coefficients $>6000$ produce a fractional gain difference which is vanishingly small, so they add little to the topological pattern of the Cryobeam. There are many modes in-between which all produce very nearly the same Cryobeam, as shown by the plateauing of each antenna in Figure~\ref{fig:trunc-test} from roughly Cryo-mode $\sim 100$ to $\sim 6000$. Therefore, it may be possible to truncate the number of Cryo-coefficients for a given beam pattern to within $\sim 100$ and fit for each of these Cryo-coefficients as model parameters.

The second extension involves using \texttt{MEDEA} to generate a training set of thousands of beams sampled uniformly from the beam hyper-parameter grid.~These beams can then be convolved with a training set of foreground maps created using the methods in \cite{hibbard_modelling_2020, hibbard_fitting_2023}, rendering a beam-weighted foreground training set which can be turned into a linear systematic model using singular value decomposition (SVD) as described in the literature \citep[for further information on using SVD to generate linear models and employing them in fits, see e.g.,][and for an implementation of this method, see the \href{https://github.com/CU-NESS/pylinex}{pylinex} software]{tauscher_global_2018, rapetti_global_2020, saxena_sky-averaged_2022}. The primary advantages of this method are the ability to readily combine sky map observations with beam simulations to obtain an efficient linear beam-weighted foreground model and the fact that the true beam model need not be within the training set. As long as the training set is accurate enough \citep[for methods to determine training set accuracy, see][]{bassett_ensuring_2021}, the remaining uncertainties within the linear model can be accounted for by the SVD coefficients, which are analytically marginalized over in the final step of the fit. Within this method, the linear beam model from \texttt{MEDEA} would serve to provide shapes for the formation of the training set that are characteristic of the beam used in the experiment; these shapes would then be compressed into the SVD modes that learn and remove them in the fitting procedure.~Ideally, this linear beam model would be constructed from a vector, $\boldsymbol{\zeta}$, of beam hyper-parameters that capture all the variations of the environment and antenna structure which sufficiently affect the shape of the beam.

As noted above, the third extension involves merely increasing the dimensionality of the input grid of beam simulations by perturbing them in other beam hyper-parameters, such as the soil $\tan \delta$, or other geometric and electrical quantities. \texttt{MEDEA} is capable of interpolating using multi-dimensional splines between beam coefficients defined on a multi-dimensional rectangular grid.

As a final note on extensions to the linear beam model, we speculate that the \texttt{Cryofunk} basis itself can be potentially used to model yet other systematics, such as the intrinsic foreground, intrinsic polarization, or any feature that can be modeled as spatially varying on a sphere. We leave such investigations to future work.

\subsection{$S$-parameter Fitting}
\begin{figure}[!t]
\centering
  \includegraphics[width=1\columnwidth]{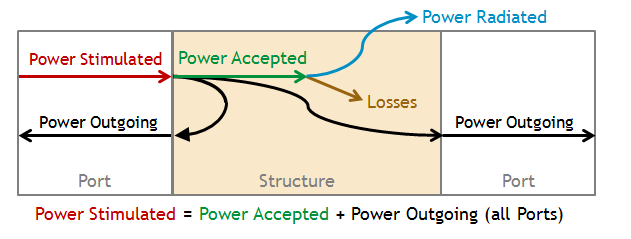} 
  \caption{Illustration of the power flow in and out for a two-port device simulation in CST, reproduced from the CST's user help guide.~For a single-port device, as is the case for all the antennas considered in this study, the extra port on the right is terminated without outgoing power, with only reflected outgoing power from the input port on the left.}\label{fig:cst_power_def}
\end{figure}

The antenna reflection coefficient, $\Gamma_\mathrm{ant}(\nu)$, describes the amount of reflected power to and from an antenna at the antenna input port \citep{rogers_absolute_2012}. This quantity $\Gamma_\mathrm{ant}(\nu) \equiv S^2_{11}(\nu)$ is commonly estimated by the single-port scattering, or $S$-parameters, measured at the antenna ports using a Vector Network Analyzer (VNA), where the subscript of $S_{11}$ refers to the specific case when input and output ports are both port \#1. This is one of the few direct measurements that can be related to the farfield beam pattern. If we denote the non-normalized radiation intensity pattern of an antenna as $U(\nu, \Omega)$, which is proportional to both the directivity $D$ and gain $G$, the total power radiated is equal to $P_\mathrm{rad} = \int_{4\pi} U(\nu, \Omega) \diff\Omega$ \citep{balanis_antenna_2005}. In CST, for a port which receives a constant stimulated power $P_\mathrm{stim}$ that is uniform across the band, the reflection coefficient can be calculated as 
\begin{equation}
    |S_{11}|^2(\nu) = 1 - \frac{P_\mathrm{rad}(\nu) + P_\mathrm{loss}(\nu)}{P_\mathrm{stim}}
    \label{eqn:s11}
\end{equation}
where $P_\mathrm{loss}$ is the power loss to the system which is not radiated. In much the same manner as for the beam-weighted foreground, the reflection coefficient is obtained by integrating the entire beam pattern and collapsing it into a `reflection coefficient' spectrum. An illustration of how the power flows through the antenna model in CST is shown in Figure~\ref{fig:cst_power_def}.

Again, the reflection coefficient can be measured, in this case in the lab, used as a data vector in a likelihood, and then fit with the same linear beam model which is being used to fit the corresponding beam-weighted foreground spectra. More powerful still would be to use a joint likelihood $\ln{\mathcal{L}_\mathrm{joint}}$ which has one component for the foreground likelihood $\ln{\mathcal{L}_\mathrm{fg}}$ and one for the reflection coefficient $\ln{\mathcal{L}_\mathrm{rc}}$, such as $\ln{\mathcal{L}_\mathrm{joint}} = \ln{\mathcal{L}_\mathrm{fg}} + \ln{\mathcal{L}_\mathrm{rc}}$. 

\begin{figure}
    \centering
    \includegraphics[width=0.46\textwidth]{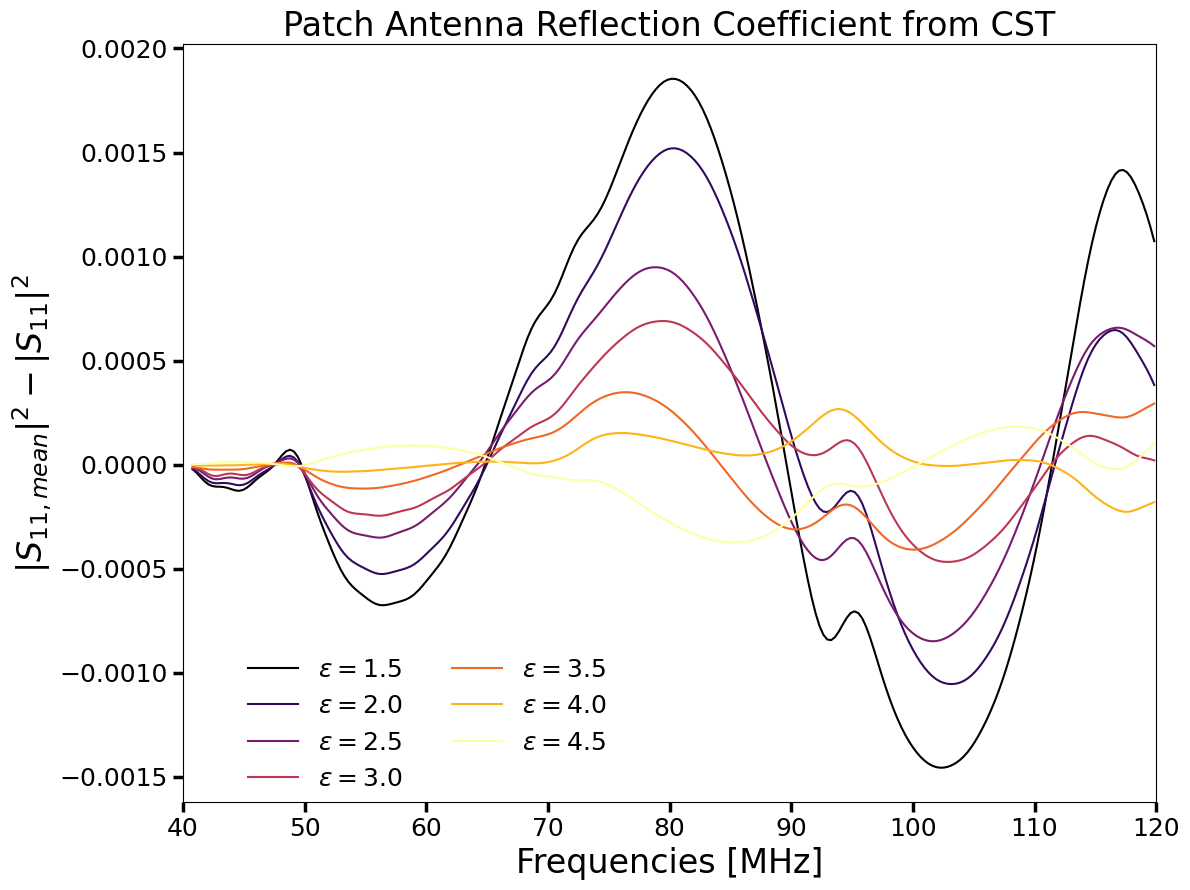}
    \caption{Relative errors between the mean and each of the Patch Antenna's $|S_{11}|^2$ values generated by CST. Note that the differences are up to $\pm 10^{-4}-10^{-2}$ depending on the frequency.}
    \label{fig:patch-s11}
\end{figure}

Figure~\ref{fig:patch-s11} shows the reflection coefficients generated by the CST simulations for the Patch Antenna as subtracted from the mean reflection coefficient at $\epsilon = 3$ in order to highlight the differences between different regolith dielectric constants. The differences are on the order of 0.0001, which is the error level reported for reflection coefficient lab measurements in, for instance, \cite{monsalve_calibration_2017}. Using Equation~\eqref{eqn:s11} for $S_{11}$, a model for $P_\mathrm{loss}$, with $P_\mathrm{rad}$ taken from \texttt{MEDEA}, and a noise level of $0.0001$, it is possible that a fit similar to those run for the foreground spectra could be run on VNA-measured reflection coefficient data to extract the regolith dielectric constant. Note that details of the radio-frequency (RF) circuit around the feed port will need to be incorporated in the CEM model to capture realistic parasitic effects on the impedance match in order to reproduce $S$-parameters closer to reality.

\subsection{Back Lobes and Soil Emission/Absorption}
The scrupulous reader may be concerned that the linear beam model presented here does not account for the possibility of back lobes present in the antenna due to its definition including only features above the horizon. Indeed, given the infinite ground regolith boundary conditions adopted for the three antenna cases, back lobes are negligible in the farfield regime. However, it is possible to instead use \texttt{MEDEA} to generate topological patterns for the back lobes by instead decomposing the portion of the sky \textit{below} the horizon. That is, we can construct a \texttt{Cryofunk} basis \textit{for the ground itself}, $Y^{\text{ground}}_{il}$. This basis could in principle then be used to model back lobes in exactly the same manner as for the main lobes. Depending on the complexity of the back lobes, such an extension could be implemented by either modeling the entire $4\pi$ steradian sky with a single \texttt{Cryofunk} basis, in which case they would be equivalent to regular spherical harmonics indexed in \texttt{Healpix} manner, or with two different \texttt{Cryofunk} bases, varying in complexity, for each portion of the map above or below the horizon.

Such an extension could also include spatially and spectrally inhomogeneous emission and absorption due to the regolith or soil, especially for experimental setups with no ground screen or which have horizons that couple to the beam pattern. There has already been recent work quantifying the effect of `hot' horizons upon 21-cm experiments \citep{pattison_modelling_2024}, and \texttt{MEDEA} could potentially improve upon such modeling.

\section{Conclusions}
\label{sec:conclusions}
We have presented a new linear beam model emulator, \texttt{MEDEA}, which can be used to rapidly and accurately generate farfield antenna beam patterns based on a subset of CEM simulations coarsely sampled from an antenna hyper-parameter space of interest. To employ \texttt{MEDEA} one first generates a sparse, input set of beam patterns using, for example, CST with each beam in the set labeled by the values of a hyper-parameter vector. Typically around ten input beams are necessary to begin, depending on the complexity of the beam pattern. Such hyper-parameters can be any antenna characteristic either electrical or geometrical which influences the antenna's beam pattern, such as the dielectric constant of the regolith/soil, length of the dipole, position of the receiver, zenith-pointing offset, and so on. \texttt{MEDEA} then decomposes each input beam in the set into the coefficients for a complete, linear basis that describes the portion of the $4\pi$ steradian sky not blocked by the experiment's horizon. Each vector in this basis is a linear combination of spherical harmonics, and because the basis is complete, \textbf{any topological beam pattern represented as a \texttt{Healpix} map can be, up to numerical precision, fully described}. \texttt{MEDEA} then interpolates between these coefficients to obtain the \textit{interpolated} coefficients for any new hyper-parameter value of interest for which \textit{no} beam was expensively simulated for the input set. One then removes one of the input beams from the input set as the test beam, and queries \texttt{MEDEA} to generate the missing beam from interpolation. If the error between the test beam and the interpolated beam is deemed too large, one must then increase the resolution of the input grid, typically requiring a doubling of the number of input beams for regular-spaced rectangular grids, after which the same test for accuracy may be employed. This process can then be repeated until the beam error falls below a desired level.

We first test \texttt{MEDEA} for three different interpolation methods with various resolutions of the hyper-parameter grid, and on three antenna case studies: (i) a small, horizontal dipole above an infinite conducting ground plane operating in a Cosmic Dawn band of $50-100$~MHz (Analytical Dipole); (ii) a patch antenna on a lunar regolith slab with a range of dielectric constant values, operating in a Cosmic Dawn band of $60-120$~MHz (Patch Antenna); and (iii) four Stacer monopoles, modeled after those of the CLPS ROLSES telescope, operating in a Dark Ages band of $1-30$~MHz (ROLSES Dipole). The interpolation methods we test include splines of various orders and Gaussian Processes with Mat\'ern kernels in Cryo-coefficient space, as well as splines applied directly to the beam pattern gains in spherical coordinate pixels. 

Next, we employ \texttt{MEDEA} as a model to fit radio spectrometer data typical of 21-cm global cosmology experiments in order to determine to what error level an input beam hyper-parameter can be recovered. Mock spectrometer data are constructed from a galactic foreground weighted with a beam from each antenna case. We add thermal noise to the data, and then fit for galactic foreground spectral indices and the beam hyper-parameter in \texttt{MEDEA} using the nonlinear sampling code \texttt{Polychord}.

We find the following:

\begin{itemize}

    \item \texttt{MEDEA} can interpolate between beam patterns from a given input set with relative errors less than $10^{-2}$ or $10^{-3}$, depending on the interpolation scheme and the smoothness of the antenna's Cryo-coefficients. For the Analytical Dipole, 20 beam simulations with a grid spacing of $\Delta h = 0.05$~m were used, 7 beam simulations with $\Delta \epsilon = 0.5$ for the Patch Antenna, and at least 20 beam simulations for the ROLSES Dipole with grid spacing $\Delta \epsilon = 0.05$. We found that in general the splines in Cryo-coefficient space tended to produce relative beam errors smaller than or comparable to the other methods (Gaussian Processes or splines in spherical coordinate pixels). Splines in Cryo-coefficient space is the interpolation method implemented in \texttt{MEDEA}.
    
    \item Beam errors on the order of $10^{-2}$ correspond to differences in antenna temperature between the true CST beam and the \texttt{MEDEA} beam (after convolution with an intrinsic foreground model) of less than $20$~mK for the Analytical Dipole, while $10^{-3}$ is needed at least for the Patch Antenna for errors less than $50$~mK. While \texttt{MEDEA} applied to the ROLSES Dipole produced similar relative beam error levels, the brighter foreground temperatures within its lower frequency range appropriate for Dark Ages experiments produce antenna temperature differences on the order of $\sim 100$~K or higher. 
    
    \item We are able to constrain beam hyper-parameters with high precision when fitting a mock beam-weighted foreground spectrum with only thermal noise using a model from \texttt{MEDEA} to generate the beam. When a systematic bias is included in the mock spectrum fits (to simulate the more realistic case of fitting real spectrometer data with models which are initially inadequate) the extracted beam hyper-parameters exhibit a bias, as expected. There are several ways to increase the model flexibility or number of parameters within \texttt{MEDEA} to potentially account for this bias. We leave such studies for future work. Our initial fits demonstrate the viability of using \texttt{MEDEA} for extracting information about the beam and in general to model 21-cm cosmology data. Furthermore, \texttt{MEDEA} is well-suited for antenna design studies, to facilitate rapid prototyping without the need to run large numbers of expensive simulations to explore the entire beam hyper-parameter space.
    
    \item As noted above, \texttt{MEDEA} as a model for the beam could be potentially enhanced by either directly fitting Cryo-coefficients from the decomposition with sampling algorithms, or by using it to generate beams to form training sets from which to produce linear beam-weighted foreground models. Regolith emission and absorption could also be easily implemented in the method by similarly decomposing the beam patterns below or on the horizon. Lastly, \texttt{MEDEA} could be used to fit antenna reflection coefficient measurements and thus further constrain the beam pattern which generates those reflection coefficients. The latter are tied to the spatial pattern of the beam and hence constrainable by our linear beam model.
\end{itemize}

\begin{acknowledgements}
    The authors would like to especially thank Henry Gebhardt for help implementing and adapting \texttt{Cryofunk} for use in this work. The authors would also like to thank Dr. Richard F. Bradley for providing the design parameters of the Patch Antenna. The authors would also like to thank Dr. Robert J. MacDowall, along with the ROLSES collaboration, for providing the preliminary dimensions of the Stacers and lander for our simulations. This work was directly supported by the NASA Solar System Exploration Research Virtual Institute cooperative agreement 80ARC017M0006. This work was also partially supported by the Universities Space Research Association via D.R. using internal funds for research development. We also acknowledge support by NASA grant 80NSSC23K0013 and a subcontract from UC-Berkeley (NASA award 80MSFC23CA015) to the University of Colorado (subcontract 00011385) for science investigations involving the LuSEE-Night lunar far side mission (JH, JB). The National Radio Astronomy Observatory is a facility of the National Science Foundation operated under cooperative agreement by Associated Universities, Inc. B. Nhan is a Jansky Fellow of the National Radio Astronomy Observatory. This work utilized the Blanca condo computing resource at the University of Colorado Boulder. Blanca is jointly funded by computing users and the University of Colorado Boulder.
\end{acknowledgements}

\appendix

\section{Forming the ROLSES Dipole from Monopoles}
\label{app-rolses-dipole}

To form a dipole pair from two separate monopole antennas situated $180^{\circ}$ apart, we begin by writing the electric fields of each monopole antenna 1 and 2 as
\begin{equation}
    \boldsymbol{E}_m = 
    \begin{bmatrix} 
    E_{\theta m}(\theta,\phi) \\ E_{\phi m}(\theta,\phi) 
    \end{bmatrix},
\end{equation}
where $m = [1,2]$ label each monopole. Note that $m=2$ is the monopole rotated spatially by $180^{\circ}$. Defining the dipole to be the $X$-arm of the antenna, the Jones matrix is
\begin{equation}
    \boldsymbol{J} = 
    \begin{bmatrix}
    E_{\theta 1} + E_{\theta 2} e^{i\pi} & E_{\phi 1} + E_{\phi 2} e^{i\pi} \\ 0 & 0
    \end{bmatrix}.
\end{equation}

The exact beam pattern used for the ROLSES Dipole is then taken to be the $M_{00}$ component of the Mueller matrix $\boldsymbol{M}$, calculated via $\boldsymbol{M} = \boldsymbol{A} \boldsymbol{J} \otimes \boldsymbol{J^*} \boldsymbol{A}^{-1}$ where we have taken the outer product of the Jones matrix with its complex conjugate, and $\boldsymbol{A}$ is a $4 \times 4$ matrix which generates Stokes vectors from the coherency vectors:
\begin{equation}
    \boldsymbol{A} = \begin{bmatrix}
        1 & 0 & 0 & 1 \\
        1 & 0 & 0 & -1 \\
        0 & 1 & 1 & 0 \\
        0 & i & -i & 0
    \end{bmatrix}.
\end{equation}
For a complete derivation of the Mueller matrix from electric field components, see Appendix A of \cite{nhan_assessment_2019}. The $M_{00}$ component of the Mueller matrix takes the Stokes total intensity component from the sky and calculates the Stokes total intensity component seen at the antenna \citep{fujiwara_introduction_2007}. Thus, the beam pattern used for the ROLSES Dipole $B_{RD}$ is given by
\begin{equation}
    B_{RD} \equiv M_{00} = \frac{1}{2} (J_{11}J^*_{11} + J_{22}J^*_{22} + J_{12}J^*_{12} + J_{21}J^*_{21}).
\end{equation}

\section{Patch Antenna Constraints}
\label{app-patch-ant-triangle-plot}
\begin{figure*}[!htb]
    \centering
    \includegraphics[width=0.96\textwidth]{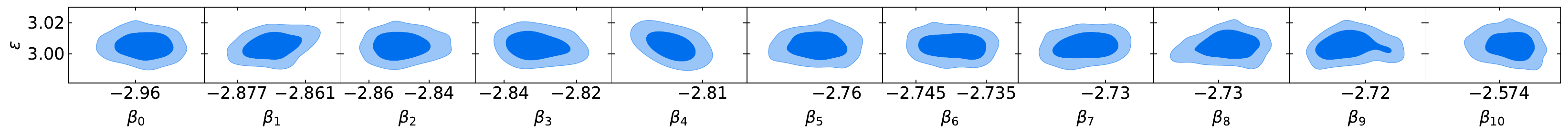}
    \caption{68 and 95 per cent confidence contours resulting from fitting our mock beam-weighted foreground spectrum to the Patch Antenna. We show only the covariances between the regolith dielectric constant, $\epsilon$, and the spectral index parameters, $\beta_j$.}
    \label{fig:patch-rectangle}
\end{figure*}

Figure~\ref{fig:patch-rectangle} shows the two dimensional posteriors resulting from fitting the Patch Antenna \texttt{MEDEA} model to our mock beam-weighted foreground spectrum. Note that only the contours which include the regolith dielectric constant $\epsilon$ are included.

\bibliography{references2}{}
\bibliographystyle{aasjournal}

\end{document}